%% file: book12_tang.tex
\documentclass[graybox]{svmult}
\pdfoutput=1

\usepackage{float}
\floatstyle{ruled}
\usepackage{graphicx}
\usepackage{amsmath}
\usepackage{verbatim}
\newfloat{program}{thp}{lop}
\floatname{program}{Listing}
\usepackage{caption}
\usepackage{multirow}
\usepackage{subfigure}

\usepackage{mathptmx}       
\usepackage{helvet}         
\usepackage{courier}        
\usepackage{type1cm}        
\usepackage{makeidx}         
\usepackage{graphicx}        
\usepackage{multicol}        
\usepackage[bottom]{footmisc}

\newcommand{\cm}[1]{}
\newcommand{\mm}[1]{}
\newcommand{\jt}[1]{}
\newcommand{\vl}[1]{}
\newcommand{\salvo}[1]{}
\newcommand{\vn}[1]{}
\newcommand{\il}[1]{}

\makeindex             

\begin{document}

\title*{Applications of Temporal Graph Metrics to Real-World Networks}
\author{John Tang\inst{1} \and Ilias Leontiadis\inst{1}Ê\and Salvatore Scellato\inst{1} \and Vincenzo Nicosia\inst{1,2} \and Cecilia Mascolo\inst{1} \and Mirco Musolesi\inst{3} \and Vito Latora\inst{2,4,5}}
\institute{
  \inst{1} Computer Laboratory, University of Cambridge 15 JJ
  Thomson Avenue, Cambridge CB3 0FD, United Kingdom.\and
  \inst{2}
Laboratorio sui Sistemi Complessi, Scuola Superiore di Catania,Via Valdisavoia 9, 95123 Catania, Italy.
  \and
  \inst{3} School of Computer Science, University of Birmingham,
  Edgbaston, Birmingham B15 2TT, United Kingdom.\and
  \inst{4} School of Mathematical Sciences, Queen Mary, University of London, E1 4NS, London, United Kingdom.
 \and 
  \inst{5}
  Dipartimento di Fisica e Astronomia and INFN, Universit\'a di Catania and INFN, Via S. Sofia 64, 95123 Catania, Italy.
}

%
%
\maketitle

\abstract*{Real world networks exhibit rich temporal information: friends are added and removed over time in online social networks; the seasons dictate the predator-prey relationship in food webs; and the propagation of a virus depends on the network of human contacts throughout the day. Recent studies have demonstrated that static network analysis is perhaps unsuitable in the study of real world network since static paths ignore time order, which, in turn, results in static shortest paths overestimating available links and underestimating their true corresponding lengths. Temporal extensions to centrality and efficiency metrics based on temporal shortest paths have also been proposed.
Firstly, we analyse the roles of key individuals of a corporate network  ranked according to temporal centrality within the context of a bankruptcy scandal; secondly, we present how such temporal metrics can be used to study the robustness of temporal networks in presence of random errors and intelligent attacks; thirdly, we study containment schemes for mobile phone malware which can spread via short range radio, similar to biological viruses; finally, we study how the temporal network structure of human interactions can be exploited to effectively immunise human populations. Through these applications we demonstrate that temporal metrics provide a more accurate and effective analysis of real-world networks compared to their static counterparts.}

\abstract{Real world networks exhibit rich temporal information: friends are added and removed over time in online social networks; the seasons dictate the predator-prey relationship in food webs; and the propagation of a virus depends on the network of human contacts throughout the day. Recent studies have demonstrated that static network analysis is perhaps unsuitable in the study of real world network since static paths ignore time order, which, in turn, results in static shortest paths overestimating available links and underestimating their true corresponding lengths. Temporal extensions to centrality and efficiency metrics based on temporal shortest paths have also been proposed.
Firstly, we analyse the roles of key individuals of a corporate network  ranked according to temporal centrality within the context of a bankruptcy scandal; secondly, we present how such temporal metrics can be used to study the robustness of temporal networks in presence of random errors and intelligent attacks; thirdly, we study containment schemes for mobile phone malware which can spread via short range radio, similar to biological viruses; finally, we study how the temporal network structure of human interactions can be exploited to effectively immunise human populations. Through these applications we demonstrate that temporal metrics provide a more accurate and effective analysis of real-world networks compared to their static counterparts.}

\section{Introduction}
\label{sec:intro}

\input{Intro}

\section{Corporate Networks}
\label{sec:corporate}
\input{Corporate}

\section{Network Robustness}
\label{sec:robustness}
\input{Robustness}

\section{Mobile Malware}
\label{sec:malware}
\input{Malware}

\section{Epidemics and Immunisation}
\label{sec:epidemics}
\input{Epidemics}

\section{Final Remarks}
\label{sec:summary}
\input{Summary}

\begin{acknowledgement}
This work was funded in part through EPSRC Project MOLTEN (EP/I017321/1) and the EU LASAGNE Project, Contract No.318132 (STREP).
\end{acknowledgement}

\bibliographystyle{plain}
\bibliography{biblio2.bib}
\end{document}

%% file: Intro.tex
Temporal graph metrics~\cite{TMML09:temporal,tang_analysing_2010} represent a powerful tool for the analysis of real-world dynamic networks, especially with respect to the aspects for which time plays a fundamental role, such as in the case of spreading of a piece of information or a disease. Indeed, existing metrics are not able to characterise the temporal structure of dynamic networks, for example in terms of centrality of nodes over time. For these reasons, new metrics have been introduced, such as temporal centrality, in order to capture the essential characteristics of time-varying graphs. A detailed description of the metrics used in this chapter can be found in~\cite{NTMMRL12:graphs}.

In this chapter we will discuss a series of possible applications of temporal graph metrics to the analysis of real-world time-varying networks. This chapter is structured as follows. We will cover our work in this area and, finally, we will discuss contributions in this fields by other researchers and potential future applications, in particular in the area of the modelling of epidemic spreading.

More specifically, in Section~\ref{sec:corporate}  we analyse the roles of key individuals according to temporal centrality within the context of the Enron scandal~\cite{tang_analysing_2010}.
In Section~\label{sec:robustness} we study how such temporal metrics can used to study the robustness of temporal networks in presence of random errors and intelligent attacks~\cite{SLMBZ12:evaluating}.
Then, Section~\ref{sec:malware} we present a containment scheme for mobile phone malware which can spread via short range radio transmission~\cite{TMML11:exploiting,THMM12:STOP}.
Finally, in Section~\ref{sec:epidemics} we discuss existing and potential applications to human epidemiology, outlining some research directions in these areas.

%% file: Corporate.tex
%

\subsection{Overview}

The Enron Energy Corporation started as a traditional gas and electrical utility supplier; however, in the late 1990s their main money making business came from trading energy on the global stock markets \cite{elkind_smartest_2004}.  
In December 2001, the Enron Energy Corporation filed for bankruptcy after it was uncovered that fraudulent accounting tricks were used to hide billions of dollars in debt \cite{_addressing2000-2001_2008}.  This led to the eventual conviction of several current and former Enron executives \cite{laurel_brubaker_calkins_enron_2004, lauren_johnston_former_2004}.  
The investigation also brought to light the reliance of the company on traders to bring in profits using aggressive tactics culminating in intentional blackouts in California in Summer 2001.  With both control over electricity plants and the ability to sell electricity over the energy markets, Enron traders artificially raised the price of electricity by shutting down power plants serving the State of California and profiting by selling electricity back at a premium \cite{cali_enron_2004}.

During the investigation into the Enron accounting scandal, telephone calls, documents and emails were subpoenaed by the U.S. government and as such the email records of 151 user mailboxes were part of the public record consisting of approximately 250,000 emails sent and received during the period between May 1999 to June 2002 (1137 days), leading up to the bankruptcy filing.  None of the emails were anonymised and so they provide unique semantic information of the owner of each mailbox. 

\subsection{Temporal Graph Construction}
 In our analysis, we use the dataset prepared by Shetty \& Adibi \cite{shetty_discovering_2005}.  Since we do not have a complete picture of the interactions of users outside of the subpoenaed mailboxes we concentrate on email exchanges between the core 151 users only.  
Taking this email dataset, we process the complete temporal graph from 1999 to 2002 with undirected links, using windows of size $w$=24 hours and horizon $h$=1.  If an email was exchanged between two individuals in a temporal window, a link between the two nodes representing those individuals will be added to the graph representing the temporal snapshot for that time. 
%
%
\begin{figure}[t]
	\begin{minipage}[h!]{0.48\linewidth} 
		\centering
		\includegraphics[scale=0.18, trim = 0mm 0mm 5mm 0mm, clip]{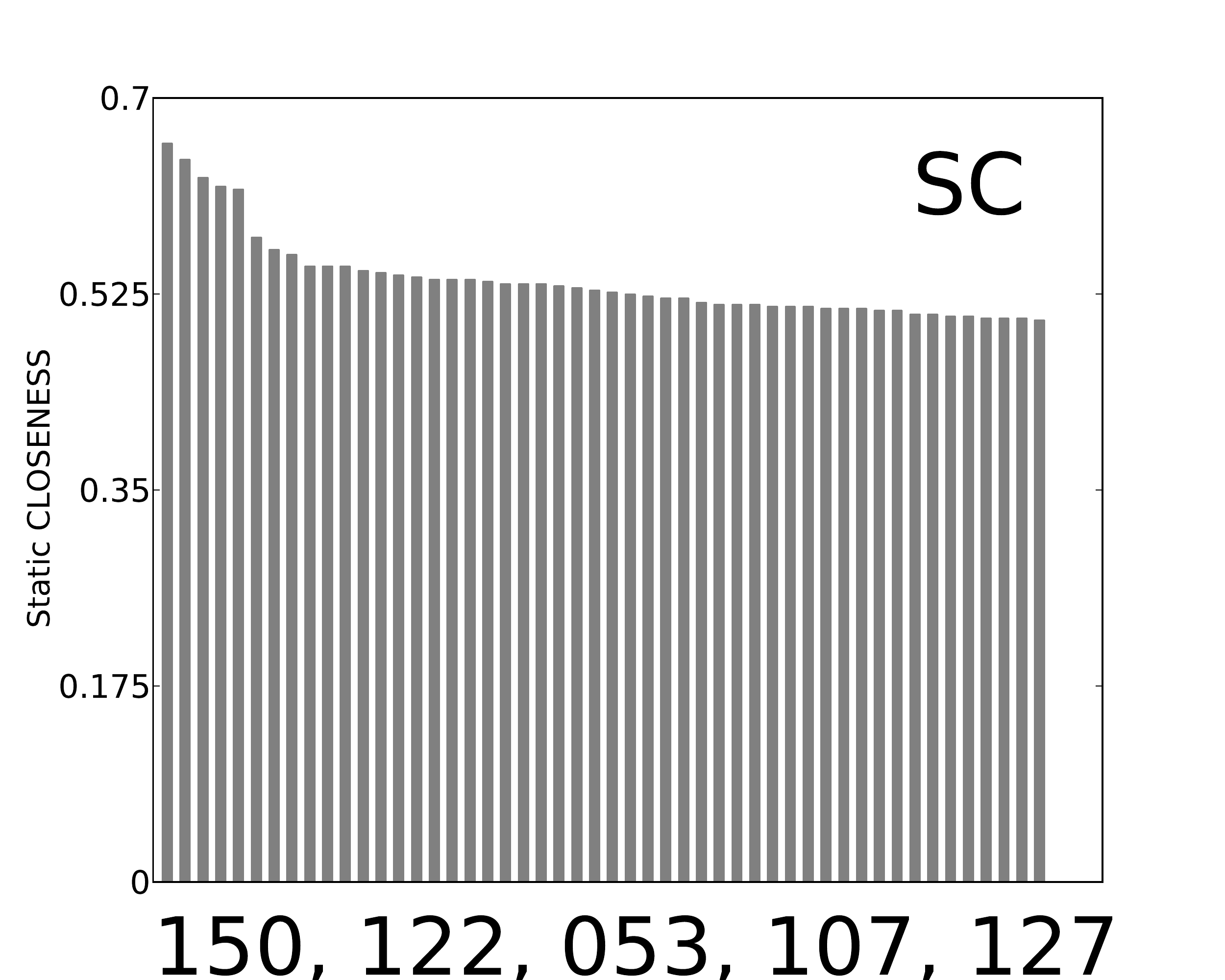}
	\end{minipage}
	\begin{minipage}[h!]{0.48\linewidth}
		\centering
		\includegraphics[scale=0.18, trim = 0mm 0mm 5mm 0mm, clip]{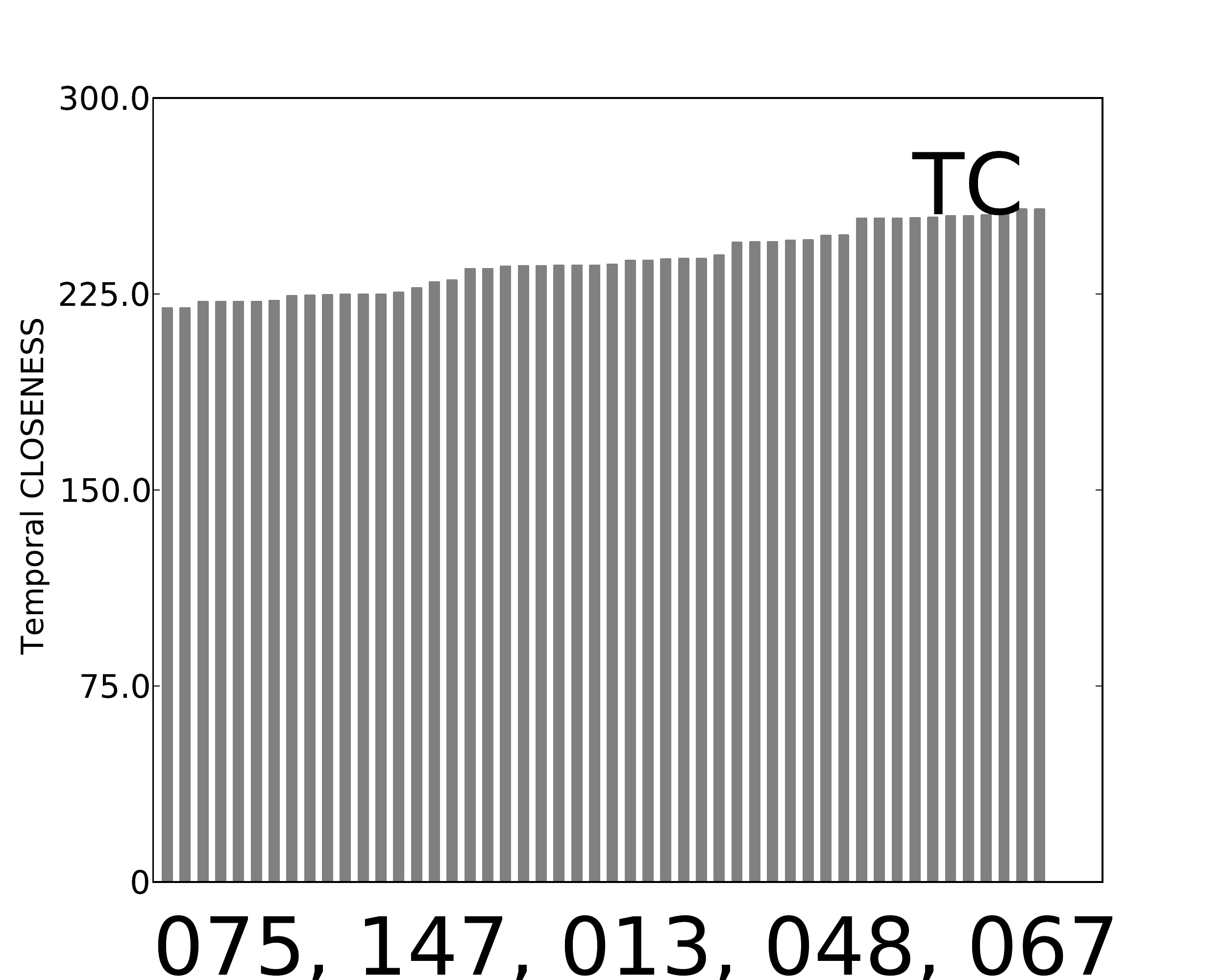}
	\end{minipage}
	
	\begin{minipage}[h!]{0.48\linewidth} 
		\centering
		\includegraphics[scale=0.18, trim = 0mm 0mm 0mm 0mm, clip]{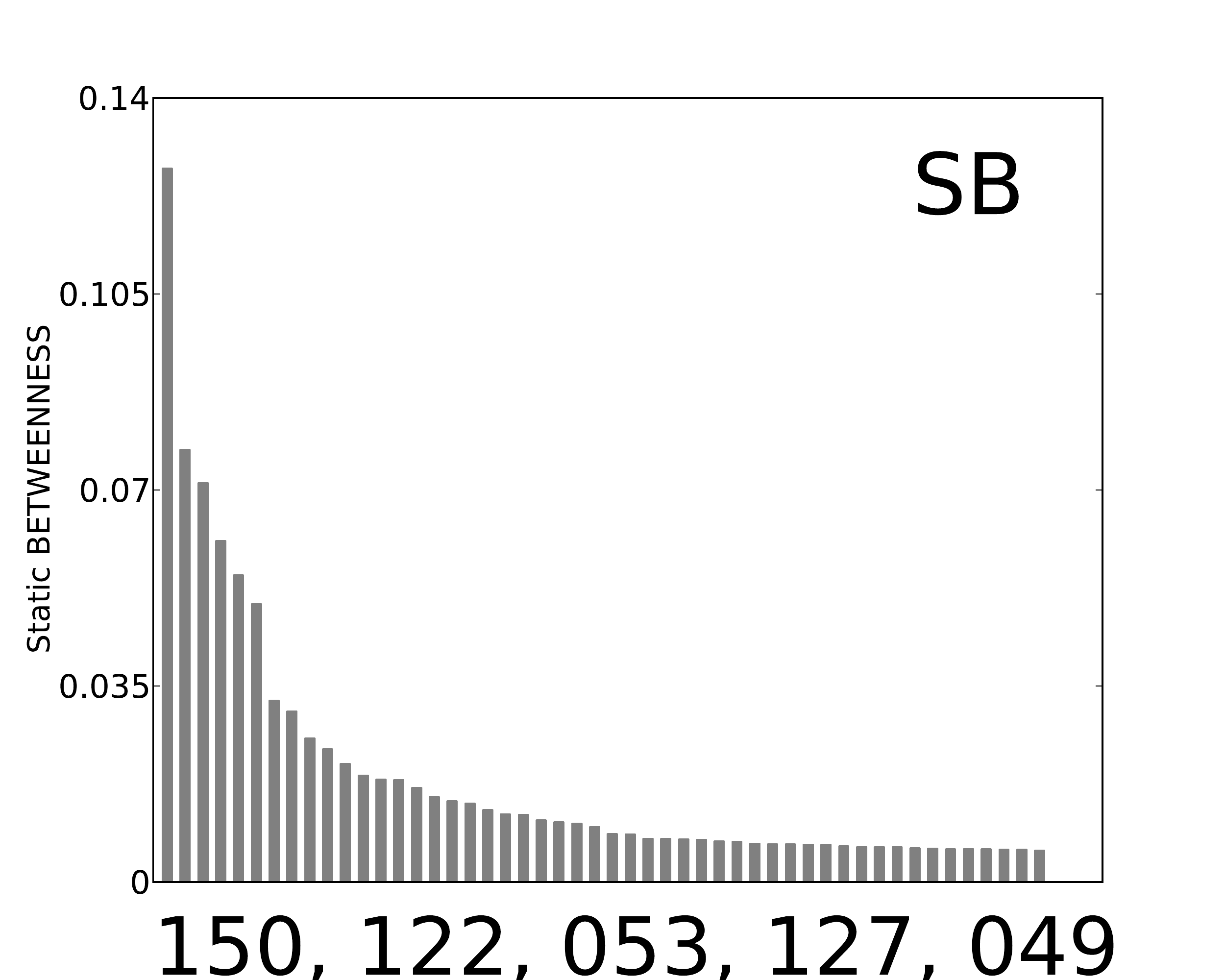}
	\end{minipage}
	\begin{minipage}[h!]{0.48\linewidth}
		\centering
		\includegraphics[scale=0.18, trim = 0mm 0mm 0mm 0mm, clip]{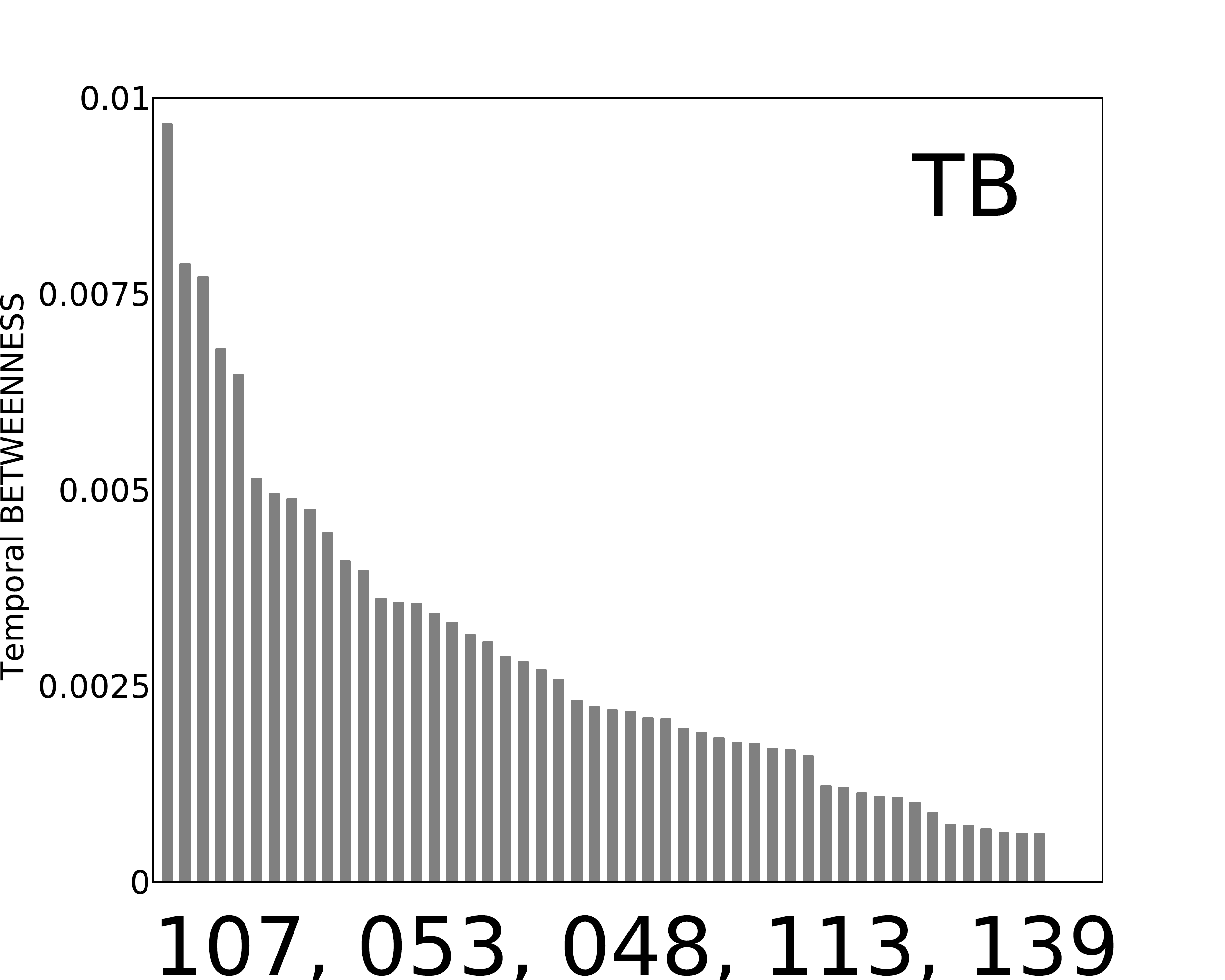}
	\end{minipage}
	
	\begin{minipage}[h!]{0.48\linewidth} 
		\centering
		\includegraphics[scale=0.18, trim = 0mm 0mm 0mm 0mm, clip]{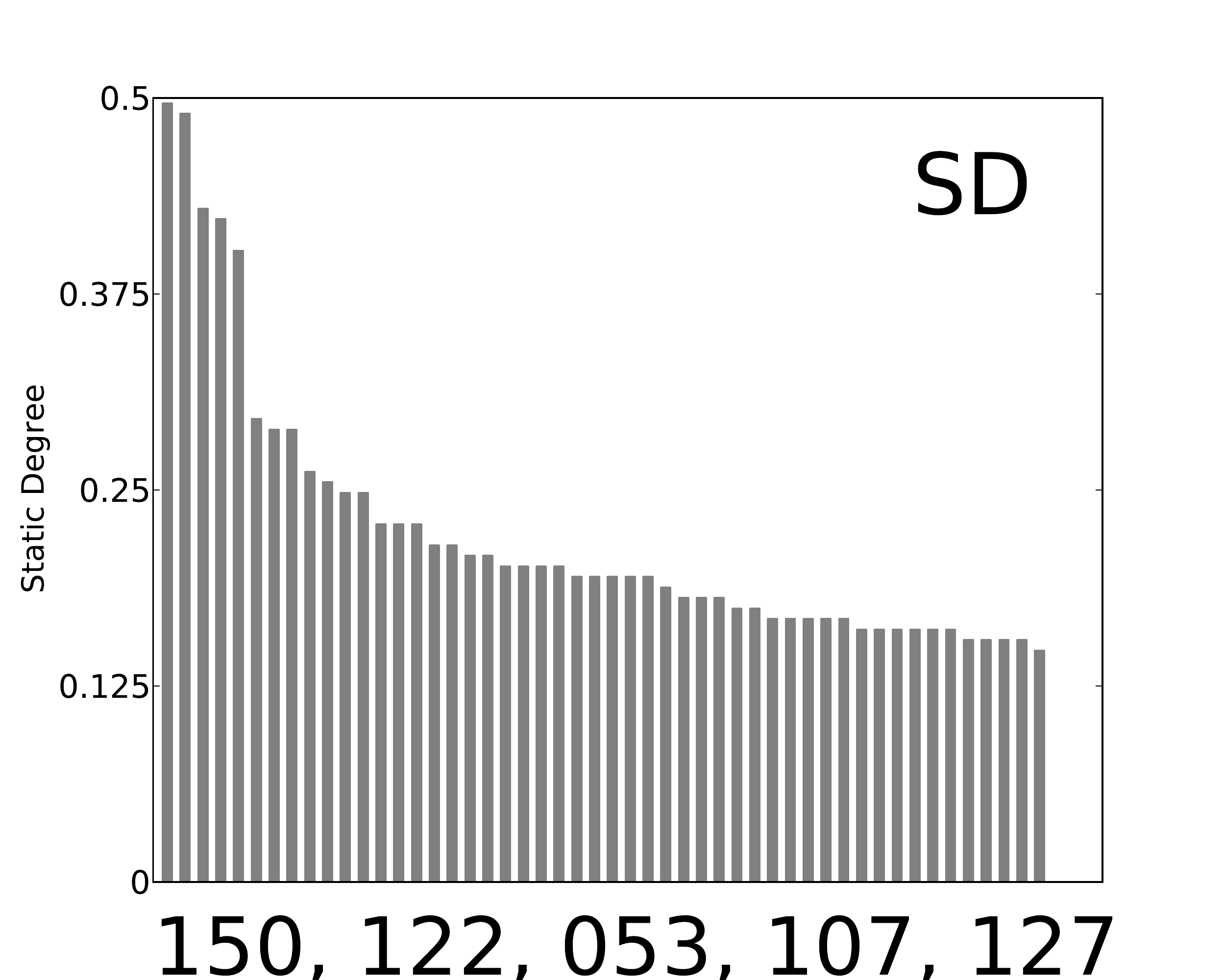}
	\end{minipage}
	\begin{minipage}[h!]{0.48\linewidth}
		\centering
		\includegraphics[scale=0.18, trim = 0mm 0mm 0mm 0mm, clip]{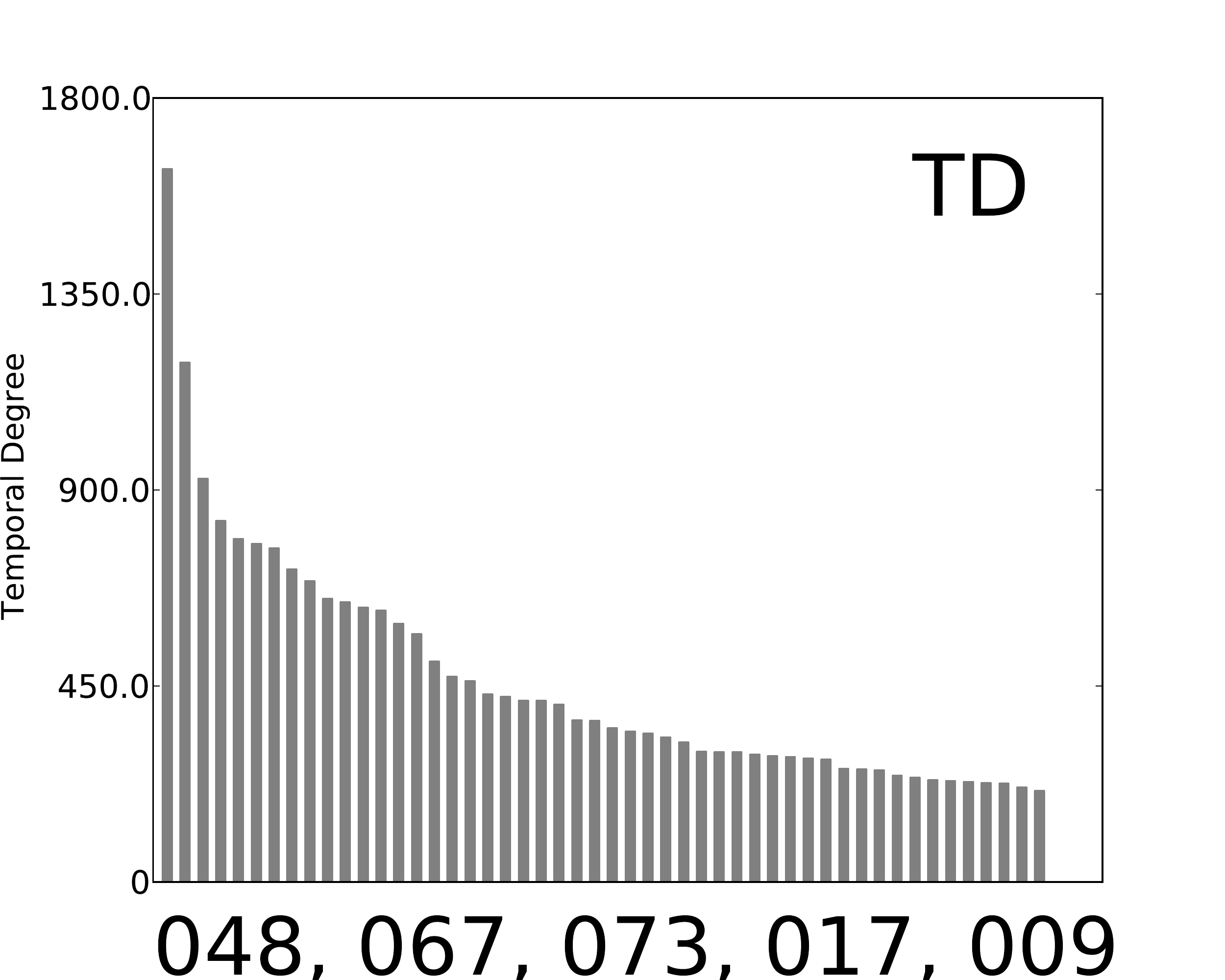}
	\end{minipage}	
	
	\caption{Ranked distribution of top 50 statically (S) and temporally (T) central nodes. From top row: Closeness (C), Betweenness (B), and Degree (D). Top 5 node ID's listed under each plot.}\label{fig:centrality}
\end{figure}
\subsection{Semantic Value of Temporal Centrality}
\label{sec:eval_semantic}
Figure \ref{fig:centrality} plots the static and temporal centrality rankings of employees calculated using closeness and betweenness.  Examining the static centralities (left column) we note that there is little difference between the top five employees using static closeness or betweenness.  Also plotting the static degree centrality of each node\footnote{The static degree centrality is defined as the number of edges connected to a node $i$, normalised by the total possible neighbour nodes $(n-1)$ \cite{wasserman_social_1994}}, we notice similar rankings suggesting that static analysis only favours employees who interacted with the most number of other people.
Temporal closeness and temporal betweenness yield different rankings amongst the top five and the calculated Kendall-tau correlation coefficient\cite{kendall_new_1938} (Table \ref{tbl:kendall}) confirm that static-to-static metrics are strongly correlated ($\simeq 0.7$). 
Also note that there is low correlation ($< 0.4$) between temporal metrics and static degree demonstrating that temporal analysis is not dependent on the number of people an individual interacts with.  

Cross referencing the top two employee identifiers with their position within the organisation (Table \ref{tbl:positions}) we identify a secretary (150) and managing director (122) as central nodes for both static closeness and betweenness; however, both temporal closeness and betweenness consistently selected employees in trading roles (053, 075, 107, 147).   A secretary and a managing director are certainly important for information dissemination and central to many communication channels, as detected by static measures. 
However, instead the top trading executives are exclusively favoured by temporal analysis.


To show that temporal analysis does not simply uncover nodes with the most interactions with other people, we also plot the temporal degree (TD) calculated as the total number of emails sent and received by each node $i$. Since there is a low correlation ($< 0.4$) with temporal closeness and betweenness this shows that temporal analysis is not dependent on the number of emails sent and received by each individual.


%

\begin{table}
\center
\small
\begin{tabular}{|c c c|}
\hline ID  & Role & Notes \\ 
\hline
9  & (Unknown) &  \\
13  & Legal & Senior Legal Specialist \\
17  & Manager &  \\
48 & Executive &  \\
53  & Trader &  \\
54  & President & Former Head of Trading \\
67  & Vice President & Enron Wholesale Services \\
73  & Trader &  \\
75 & Director of Trading &  \\
107  & Trader & Head of Online Trading \\
122  & Managing Director &  \\
127  & Chairman \& CEO &  \\
139  & Director &  \\
147  & Trader &  \\
150  & Secretary & Assistant to Greg Whalley \\
\hline 
\end{tabular} 
\caption{Roles of top centrality nodes.} \label{tbl:positions}
\end{table}

\begin{table}
\center
\small
\begin{tabular}{|c | c c c c c c|}
 \hline
 & SB & SC & SD & TB & TC & TD \\ 
 \hline
SB & 1.00 & 0.57 & 0.69 & 0.41 & 0.24 & 0.43 \\ 
SC & - & 1.00 & 0.70 & 0.36 & 0.22 & 0.31 \\ 
SD & - & - & 1.00 & 0.39 & 0.28 & 0.48 \\ 
TB & - & - & - & 1.00 & 0.43 & 0.34 \\ 
TC & - & - & - & - & 1.00 & 0.40 \\ 
TD & - & - & - & - & - & 1.00 \\ 
\hline 
\end{tabular} 
\caption{Kendall-tau correlation coefficients between centralities.} \label{tbl:kendall}
\end{table}

\begin{figure}[t]
	\begin{minipage}[h!]{0.49\linewidth} 
		\centering
		\includegraphics[scale=0.18, trim = 0mm 0mm 0mm 0mm, clip]{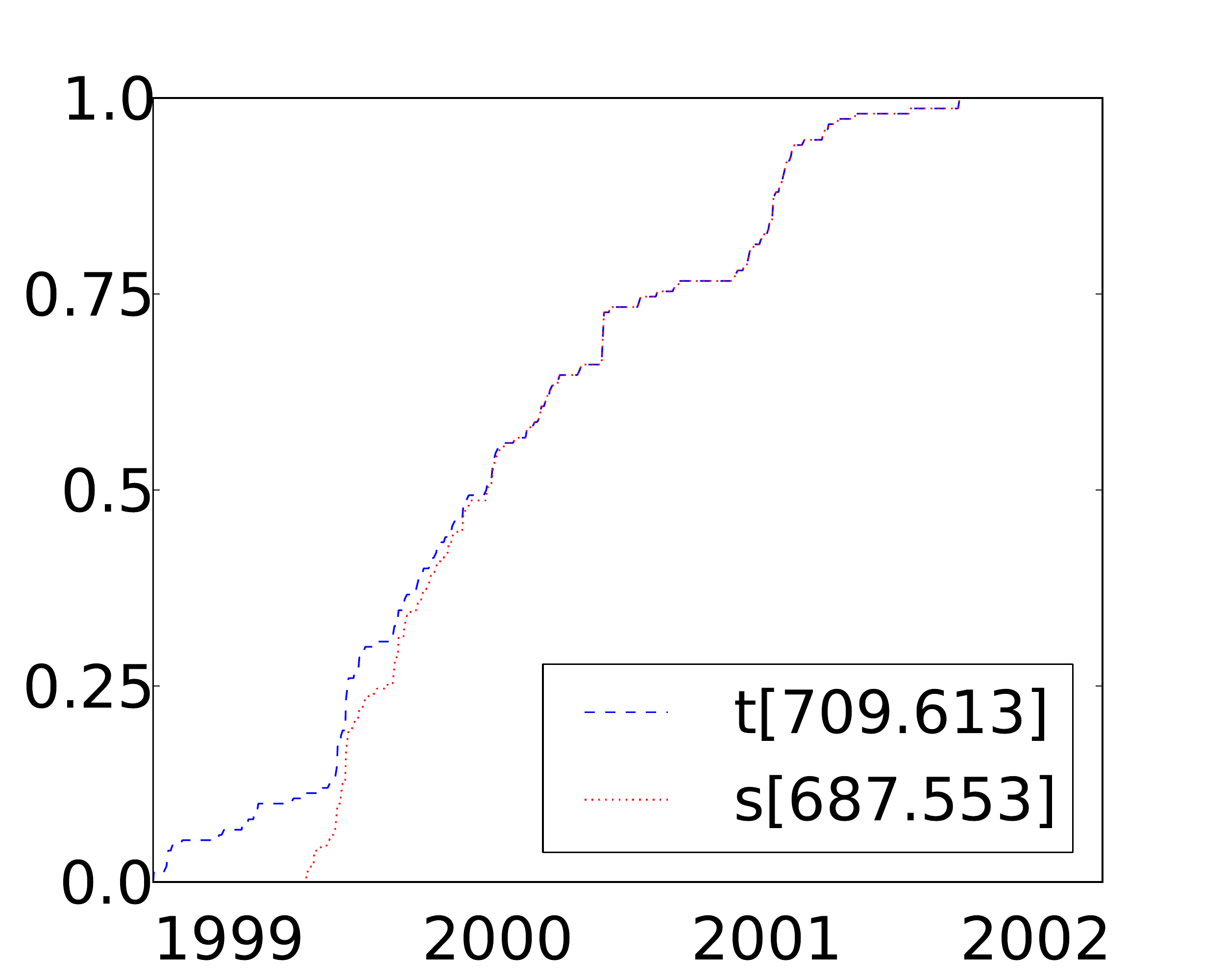}
	\end{minipage}
	\begin{minipage}[h!]{0.49\linewidth}
		\centering
		\includegraphics[scale=0.18, trim = 0mm 0mm 0mm 0mm, clip]{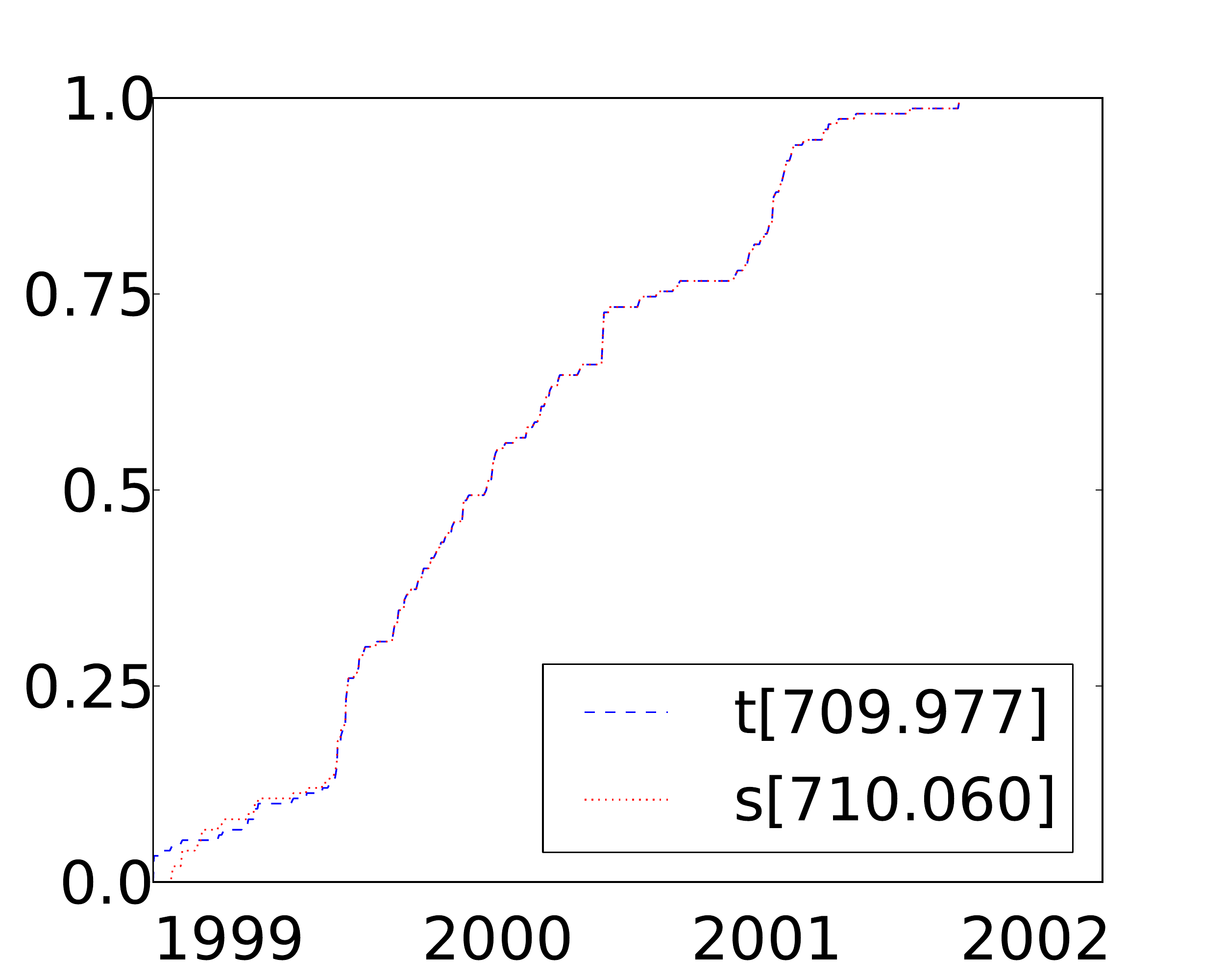}
	\end{minipage}
	\caption{Dissemination Process: Dissemination ratio starting from top 2 (left) and top 10 (right) closeness source nodes.  Area under curve reported in legend for temporal (t) and static (s) centrality.}\label{fig:dis_cl_n1_n10}
\vspace{-5pt}
\end{figure}

\medskip

\subsection{Effectiveness of Central Nodes on Dynamic Processes}
\label{sec:eval_centrality}

\subsubsection{Trace-driven Simulation Setup}
To evaluate the role and the centrality of the employees identified by temporal and static analysis, we consider two dynamic processes.  First, we simulate a simple information \textit{dissemination process} over the temporal graph constructed from the Enron traces. The process is simulated as follows.
We select the top $N$ nodes from the ranking based on temporal closeness centrality. We place an identical message $m$ into their (infinite) buffers.  We refer to any node that has received a copy of this message as \textit{reached}.  
We then replay the contact trace through time and as reached nodes
make contact with an unreached node $u$, the message is replicated into
the buffer of node $u$. We assume that messages are transferred instantaneously and only the
first neighbour in a time window can be reached.  
We then repeat this for static closeness centrality and plot the dissemination ratio across time for both.  

Second, to model the role of individuals as part of an information \textit{mediation process}, we borrow concepts from the more commonly known epidemic immunisation process where the dissemination ratio of a contagion spreading throughout a static network is measured before and after certain nodes are immunised against the contagion \cite{barrat_dynamical_2008}.  This is analogous to measuring the spread of information (the contagion) before and after important individuals are removed from the network (such as going on holiday or being discharged) since our conjecture is that removing mediators will impact the network communication efficiency greatly.  

In the trace-driven simulation, instead of a single message spreading within the organisation, we seed all employees with a different message that needs to be delivered to all other employees. This models multiple channels of communication.   In order to derive a baseline performance, we start by calculating the dissemination ratio when no nodes are removed.  We then remove the top $N$ individuals identified by temporal betweenness and rerun the information spreading process.  Nodes which are removed cannot receive or pass on messages.  We then repeat the same process for comparison using static betweenness centrality for the ranking.
%
%
%
%
\begin{figure}[t]
	\begin{minipage}[h!]{0.49\linewidth} 
		\centering
		\includegraphics[scale=0.18, trim = 0mm 0mm 0mm 0mm, clip]{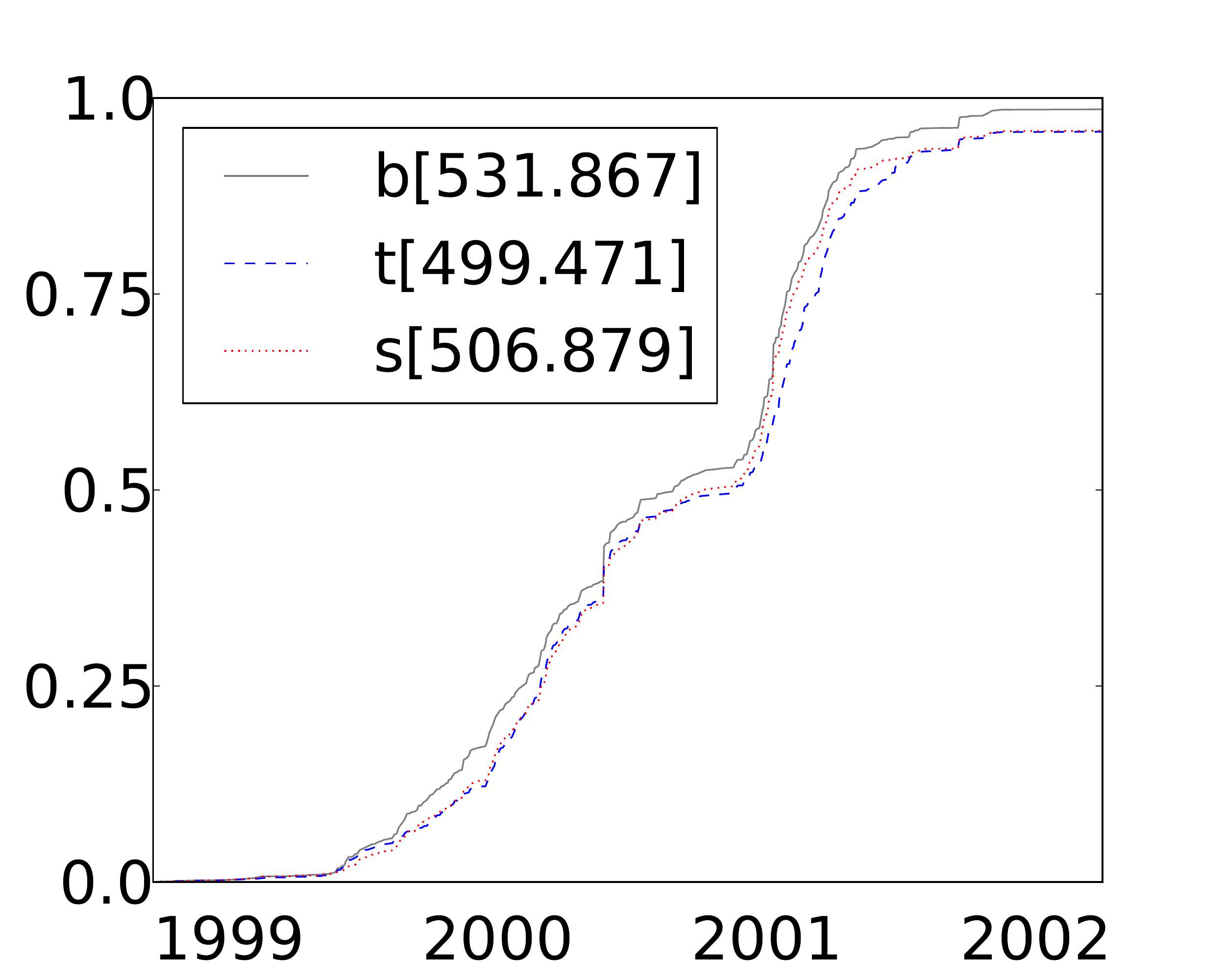}
	\end{minipage}
	\begin{minipage}[h!]{0.49\linewidth}
		\centering
		\includegraphics[scale=0.18, trim = 0mm 0mm 0mm 0mm, clip]{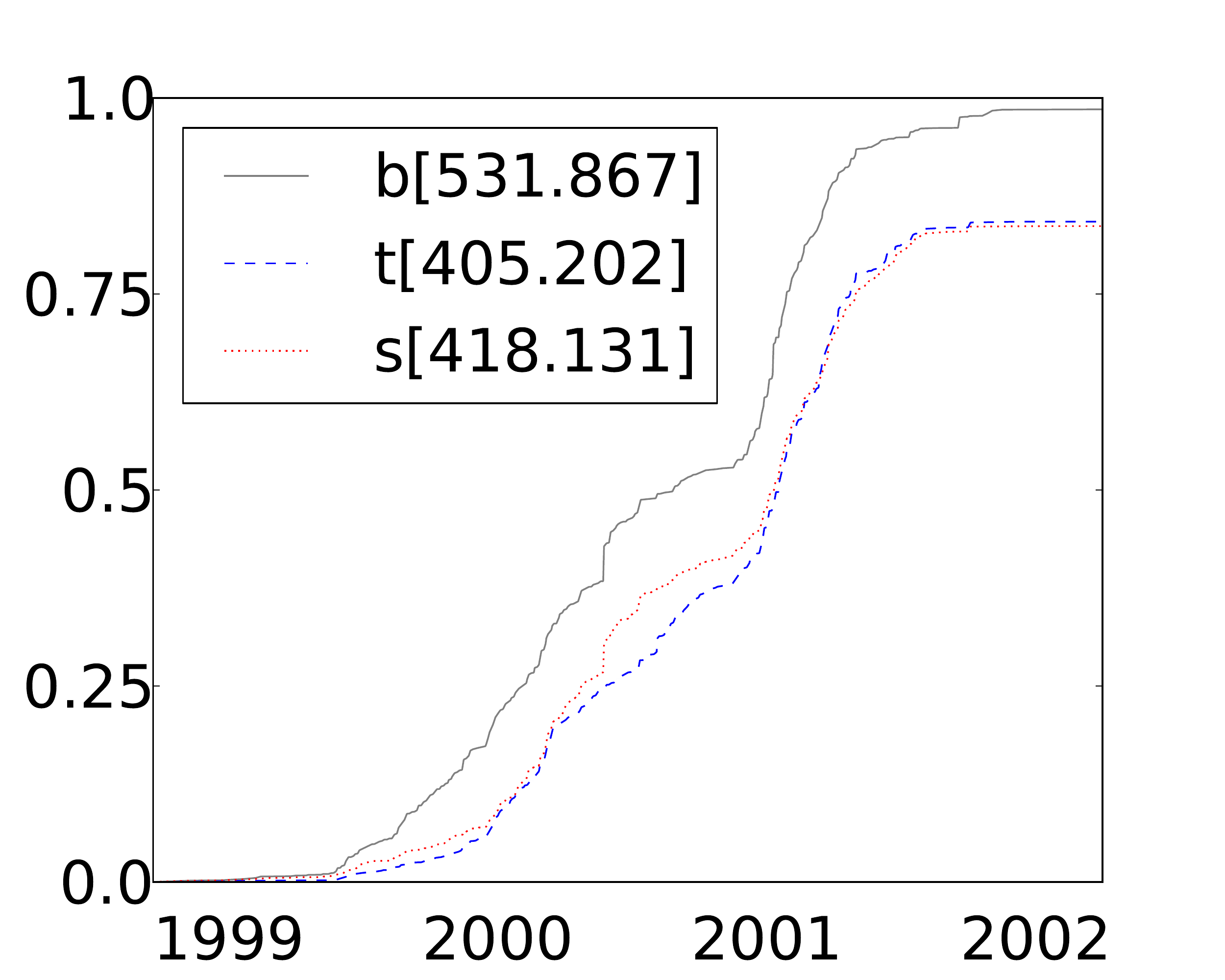}
	\end{minipage}
	\caption{Mediation Process: Dissemination ratio after removing top 2 (left) and top 10 (right) betweenness nodes. Area under curve reported in legend for temporal (t), static (s) and baseline (b) where no nodes are removed.}\label{fig:ep_bet_n1_n10}
\end{figure}
%
%
\subsubsection{Evaluating Information Dissemination \& Mediation}
We present plots using $N=\{2,10\}$ for information dissemination (Figure \ref{fig:dis_cl_n1_n10}) and information mediation (Figure \ref{fig:ep_bet_n1_n10}).  As we can see the different pairs of traders identified by temporal analysis are better than the arbitrary nodes selected by static analysis for both disseminating information through the organisation and acting as mediators between communication channels.  In the information dissemination case, although the final dissemination is the same across the long period of time, the two traders selected by temporal analysis disseminate information quicker.  Only after increasing to 10 nodes the static analysis presents similar results.  
In the information mediation case, the final dissemination ratios for both temporal and static centrality nodes slightly decreases by removing the nodes but are comparable.
However, removing the two traders gives an overall more prolonged drop in message dissemination.  In the case of the removal of 10 nodes, the individuals identified by means of the temporal metrics slow the dissemination process further compared to static ones.

\subsection{Summary of the Findings}
This study has demonstrated the advantages of using temporal network analysis over traditional static, aggregated graphs.  Although centrality rankings are quantitative measures of node importance, their physical meaning is very much qualitative.  For this reason, we have shown that temporal centrality provides a more accurate identification of key people in a corporate social network where each persons role is known.  Taking a second perspective, we demonstrate that temporal centrality can identify nodes which can spread and mediate information, better than static analysis.  This demonstrates the importance of temporal information for applications which are applied to real networks.

%% file: Robustness.tex
\subsection{Overview}
The study of real-world communication systems by means of complex network models
has provided insightful results and has vastly expanded our knowledge on how
single entities create connections  and how these connections are used for
communication or, more generally, interaction~\cite{BOCCALETTI2006}. 
In particular, technological networks such as the Internet and the World Wide Web
have been under scrutiny  in terms of structure and dynamic
behavior~\cite{Huberman99,Faloutsos1999}. More recently, with the
widespread adoption of mobile and opportunistic networks, it has
become important to develop new analytical tools to keep into
account network dynamics over time~\cite{Kostakos2009,Kempe2000}
and how this affects phenomena such as information
propagation~\cite{CE07:persistence, Chaintreau_CHCDGS07:Impact}.

At the same time, the problem of understanding whether real systems can sustain
substantial damage and still maintain acceptable performance  has been
extensively addressed~\cite{albert_error_2000,callaway2000}.  Various measures of network robustness
have been defined and investigated for several classes of networks, evaluating
how different system can be more or less resilient against random errors or
targeted attacks thanks to their underlying structural
properties~\cite{holme2002,crucitti2004}. 

Nonetheless, it is still unclear how to approach the study of robustness of
networks by taking into account their time-varying nature: by adopting a static
representation of a temporal network,  important features which impact the
actual performance might be missed. Thus, it becomes important to develop a
robustness metric which takes into account the temporal dimension and gives
insights on how a mobile network is affected by damage or change.  Particularly,
the fact that links are not always active means that information
spreading can be delayed or even stopped and that {\em relative ordering in
time of connection events may affect the creation of temporal paths in
mobile networks}. 

Our main goal is to design a novel framework for the analysis of robustness in
time-varying networks. We adopt \textit{temporal network
metrics}~\cite{TMML09:temporal} to quantify network performance and define a
measure of robustness against generic network damages. At first, we study its
performance on random network models to understand its properties; then we
apply our method to study a real mobile network, describing how temporal
robustness gives a more accurate evaluation of system resilience than static
approaches.  

\subsection{A Framework to Evaluate Robustness of Temporal Networks}
The study of robustness of complex networks has mainly focused on describing how
a given performance metric of the network is affected when nodes are removed
according to a certain rule. The underlying assumption is that the absence or
malfunctioning of some nodes will cause the removal of their edges and, thus,
some paths will become longer, increasing the distances between
the remaining nodes, or completely disappear, resulting in the
loss of connectivity in the whole system.  In this work we will
study the problem of defining and analyzing robustness in
evolving networks: as a consequence, {\em we need to use a
performance metric which includes the temporal dimension in its
definition}.  We choose to adopt temporal efficiency as the
performance metric.  We then consider random and independent
failures for every node and we evaluate how the system
tolerates increasing level of malfunctioning nodes.

Since a temporal graph is continuously evolving, we can evaluate how
temporal efficiency changes over time by considering a value $\tau$ and
evaluating $E_G(t)$ as the relative temporal efficiency of the temporal graph in
the time window $[t-\tau,t]$. The effect of $\tau$ is to effectively impose an
upper bound on the temporal distances, as all paths longer than $\tau$ simply do
not exist. As a consequence, $\tau$ should be chosen so that any communication
whose delay is longer than $\tau$ itself can be ignored. 

Given a temporal graph $G$, we define a damage $D$ as any structural
modification on it and we define $G_D$ as the graph resulting by the effect of
the damage $D$ on $G$. A damage may be the deactivation of some nodes or the
removal of some edges at a particular time $t_D$. Because of damage $D$, some
temporal shortest paths will be longer or will not exist any more, thus, we
expect that the temporal efficiency will eventually reach a new steady value
$E_{G_D} \leq E_G$. It is important to evaluate the new value of the temporal
efficiency on a new temporal graph which still contains the deactivated nodes,
in order to obtain a decrease in efficiency. Otherwise, we might
obtain a smaller temporal graph which is more efficient than the
original graph, although it has lost much of its structure. 
Hence, we do not consider highly dynamic systems where nodes can be
constantly added or removed. Instead, we focus on evaluating the service
degradation in a more controlled environment where only a number of existing nodes
could fail.

We define the loss in efficiency $\Delta E(G,D)$ caused by the damage $D$ on the
temporal graph $G$ as $\Delta E(G,D) = E_G - E_{G_D}$.  Finally, we define the
\textit{temporal robustness} $R_G(D)$ of the temporal graph against the damage $D$ as
\begin{equation}
\label{eq:robustness}
R_G(D) = 1 - \frac{\Delta E(G,D)}{E_G} = \frac{E_{G_D}}{E_G}
\end{equation}
This value is normalized between 0 and 1 and it measures the relative loss of
efficiency caused by the damage: if the damage does not impact the efficiency of
the graph ($E_{G_D} = E_G$) then its robustness is 1, while if the damage
destroys the efficiency of the graph ($E_{G_D} = 0$) the robustness drops to 0.
Temporal efficiency is a particularly suitable metric to study temporal network
robustness as it denotes both longer temporal paths and the lack of paths among
temporally disconnected nodes at the same time. Nonetheless, other metrics have
been used to assess robustness in static systems: for instance, there could be
scenarios where fast communication with small delays can be more important than
global connectivity, thus other measures can be adopted. Provided that these
measures can be
extended to the temporal case, they can be easily integrated in our framework.


\subsection{Robustness of Random Temporal Networks}
In this section, we present a numerical analysis of temporal 
robustness for different classes of random temporal networks:
an Erd\H{o}s-R\'enyi temporal model, a Markovian temporal model and mobility-based temporal model.

\subsubsection{Random Temporal Network Models}
An Erd\H{o}s-R\'enyi (ER) random graph with $N$ nodes and parameter $p$  
is created by independently including each possible edge in the graph with
probability $p$ and it is denoted as $G(N,p)$~\cite{ErdRen60}.  
We generalize this model to the temporal case by creating a sequence of $T$ ER
random graphs $G(N,p)$ and we denote the resulting temporal graph as $G(N,p,T)$.

The temporal ER network model does not provide temporal correlations between
consecutive graphs in the sequence. We thus consider a model where link evolution
is described by a Markov process, thus enabling memory effects in network
dynamics. We consider a complete graph $G$ with $N$ nodes. At every discrete timestep $t$
each link may or may not be present: a temporal graph is created
where the existence of each link evolves according to a 2-state discrete Markov
process. We denote with $p$ the probability that a link present at time $t$ will be
removed at time $t+1$ and with $q$ the probability that a link which is not
present at time $t$ will be added at time $t+1$.  The steady probability of link
presence then is $P_{ON} = \frac{q}{p+q}$: as a consequence, each observation of
the temporal graph appears as an ER random graph with each edge present with
probability $P_{ON}$. 

We also create a random model of a temporal network by using mobility
models. In this case we are introducing 
topological constraints: a key difference with the previous temporal
models is that each node is not equally likely to connect with all the other
nodes, due to the effect of spatial distance. 
We consider $N=100$ nodes moving in a square area $1000$x$1000$ meters and we
define a communication range $r$: at every time step, we create a 
graph where nodes are connected if their Euclidean
distance is shorter than $r$. Thus, we change the probability of link
presence $P_{ON}$ by varying the communication range.
Then, a temporal graph can be defined as the sequence of graphs extracted at
each time step while the nodes move. We investigate two different mobility models that are implemented using the
Universal Mobility Model Framework~\cite{ummf}: Random Waypoint Model (RWP) and
Random Waypoint Group Model (RWPG). In RWP each node selects uniformly at random a location towards which it moves 
with speed uniformly distributed in a fixed range
$[5, 40]$ mph. As the node reaches its destination, it waits for a
randomly distributed time in $[0, 120]$ seconds and repeats the above steps
until the end of the simulation. 

In RWPG nodes are divided into two classes: there are $M$ group leaders and $N-M$
group followers. Every group followers has its own leader so that the $N$
nodes are divided into equally-sized groups.  Each group leader selects a random
target and moves towards it, according to the RWP mobility model. Group members
do not select any target; instead, they follow their group leader according to
the \emph{pursuit force}~\cite{ummf} which is set to give a group span of 200
meters.

\subsubsection{Numerical Evaluation}
We numerically evaluate temporal efficiency $E_G(t)$ over time, adopting a time
window of $\tau=100$, for a graph with $N=100$ nodes: after an initial phase,
the random temporal graph reaches an equilibrium state and we compute the
steady value of temporal efficiency. We run each simulation for $2\tau$
steps and we compute the average value of temporal efficiency over the
last $\tau$ steps. All results have been averaged over 100 different
runs.  We evaluated numerically temporal robustness
by deactivating each node independently with
increasing probability $P_{ERR}$. We measure temporal efficiency before and after the
failure, when the network reaches a new equilibrium state.

\begin{figure}[th]
  \centering
  \subfigure[$N=100$]
      {\label{fig:er_robustness_100}
          \includegraphics[width=0.48\columnwidth]{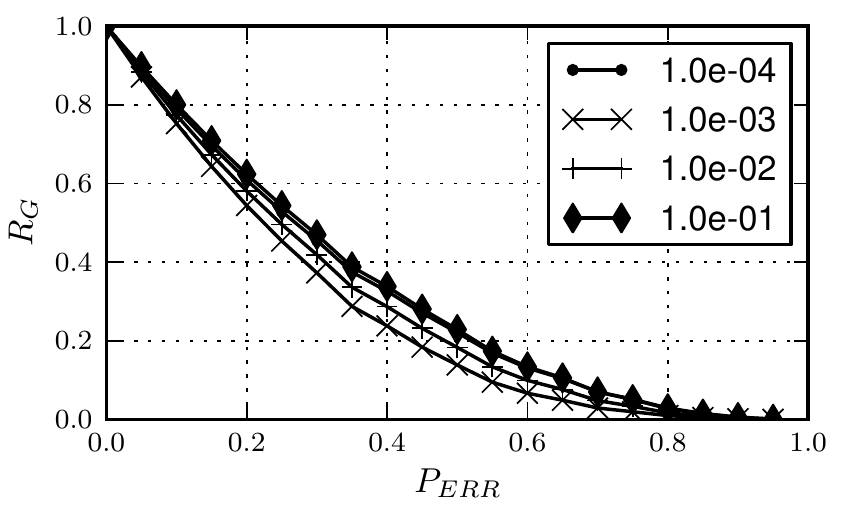}}
  \subfigure[$N=1000$]
      {\label{fig:er_robustness_1000}
      \includegraphics[width=0.48\columnwidth]{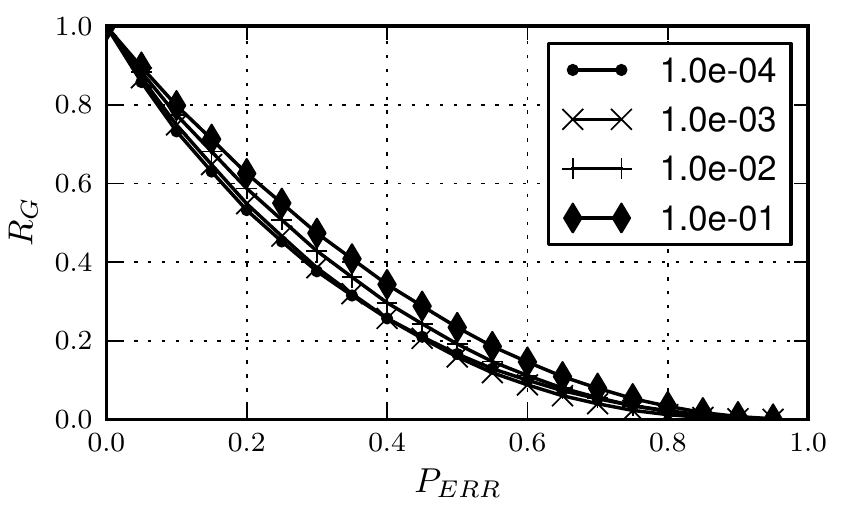}}
  \caption{Temporal robustness $R_G$ as a function of probability of error $P_{ERR}$
      in the ER random model for different link probability $p$.
          The size of the
          system has no impact on temporal robustness: furthermore, the system
          fails smoothly as the probability of error increases.}
    \label{fig:er_robustness}
\end{figure}

\begin{figure}[t]
  \subfigure[Markov]
{\label{fig:mobility_robustness_markov}
    \includegraphics[width=0.48\columnwidth]{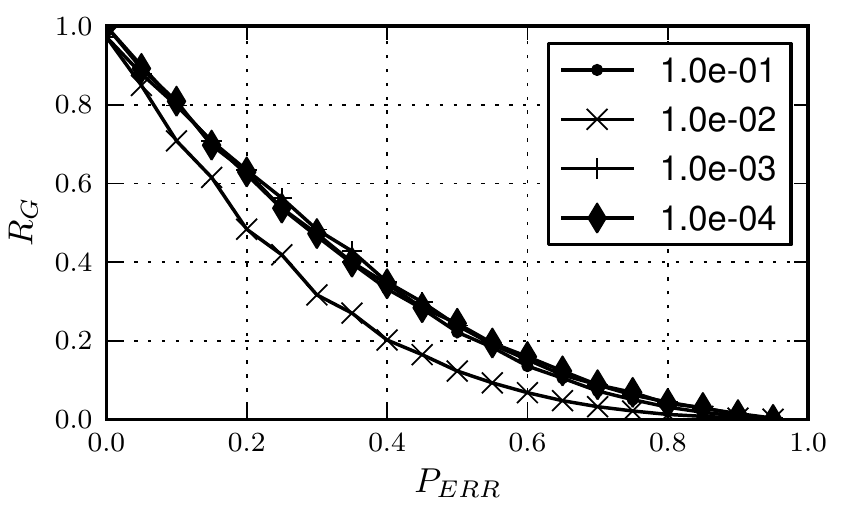}}
  \subfigure[RWP]
{\label{fig:mobility_robustness_rwp}
\includegraphics[width=0.48\columnwidth]{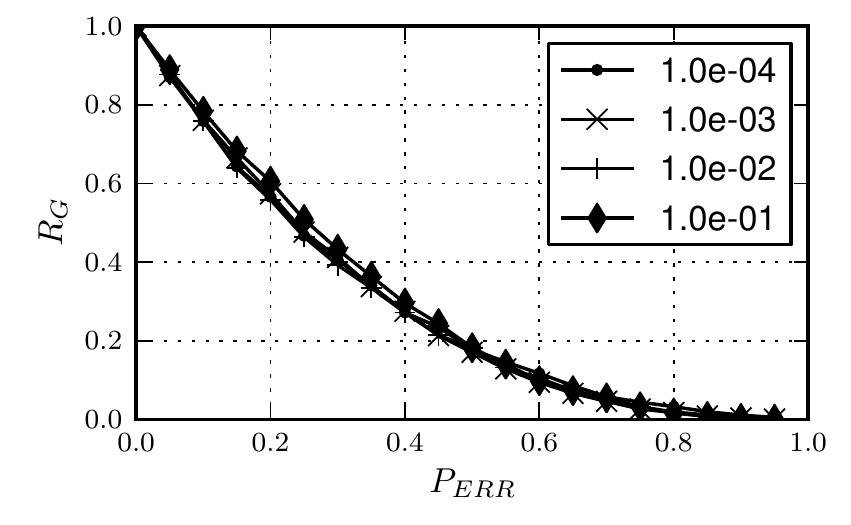}}
  \caption{
  Temporal robustness $R_G$ as a function of probability of error
      $P_{ERR}$ and for different values of $P_{ON}$ for the Markov-based and
      for the RWP random model (RWPG does not deviate from RWP)}
    \label{fig:mobility_robustness}
\end{figure}

As reported in Figure~\ref{fig:er_robustness}, the temporal ER model 
fails smoothly as we increase the fraction of removed nodes,
 without any sudden disruption for any value of $P_{ERR}$.  This is a main
 difference with respect to what happens in the static case: for a static ER
 random graph there may exist a critical value of $P_{ERR}$ which causes a breakdown of
 the network in several disconnected components~\cite{albert_error_2000}.
 This is not true for temporal robustness, as new paths can still appear
 after the damage as the network rearranges its connections. Time provides
 more redundancy and, hence, more resilience.  Moreover, we also note that
 temporal robustness does not depend on system size: since it is normalized
 with respect to the value of temporal efficiency before the damage, it
 depends only on the relative drop in efficiency, not on absolute values.

As shown in Figure~\ref{fig:mobility_robustness_markov}, temporal robustness is
affected by probability of error $P_{ERR}$ in the same way as in the
temporal ER model: the system fails gradually as more nodes are removed. However,
for intermediate values of $P_{ON}$ robustness has lower
values.  At the same
time,  high and low values of $P_{ON}$ provide the same
robustness, even if 
the absolute value of temporal
efficiency can be very different, thanks to the normalization of temporal
robustness.

In the case of mobility-based temporal networks, reported in
Figure~\ref{fig:mobility_robustness_rwp}, both RWP and RWPG exhibit a similar
behavior: again, the network loses efficiency in a smooth way and temporal
robustness is not affected by $P_{ON}$ in this case as the spatial
characteristics of the network are mainly affecting the resulting robustness.


\subsection{Case Study: Cabspotting}
We have seen that temporal networks do not exhibit sudden breakdowns when nodes are being
removed and that various temporal network models exhibit analogies in
their resilience. We now shift our attention to real time-varying networks: our
aim is to understand whether temporal robustness gives us more information
than static robustness in a real case and to investigate whether random models can
offer a good approximation to real networks.

\subsubsection{Dataset}
This case study is based on Cabspotting, a publicly available dataset of
mobility traces: the Cabspotting project tracked taxi cabs in San Francisco
traveling through all the Bay Area for about two years with the aim of gathering
data about city life~\cite{cabspotting}.  The vehicles were equipped with GPS
sensors and every device was periodically updating its position and uploading it
to a central server to be stored, along with the timestamp of the record. Thus,
   it is possible to reconstruct each taxi's trajectory over space. For pictorial representations of the dataset, please refer to the project website~\cite{cabspottingproject}.

We have selected an area of about 20 km x 20 km around the city of San Francisco
and we have extracted 24 consecutive hours of mobility traces, corresponding to
Wednesday, 21 May 2008. After this, we have generated an artificial contact
trace by defining a communication range of 200 m for the vehicles, which roughly
corresponds to WiFi connectivity range in similar
scenarios~\cite{Chaintreau2009}: whenever two cars are within this distance they
can communicate to each other. Time
granularity is in seconds, so we have a sequence of 86,400 graphs with 488
nodes and more than 350,000 contacts among them. The average contact duration
is about 2 minutes while the average inter-contact time is more than 2.5 hours.

\begin{figure}[t]
  \centering
  \subfigure
{\label{fig:cabspotting_robustness_static}
    \includegraphics[width=.48\columnwidth]{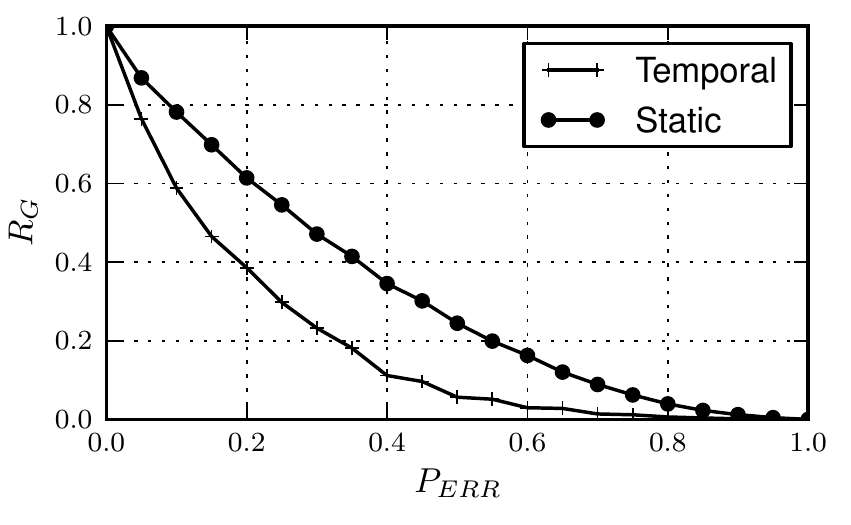}}
\subfigure
{\label{fig:cabspotting_robustness_null}
\includegraphics[width=.48\columnwidth]{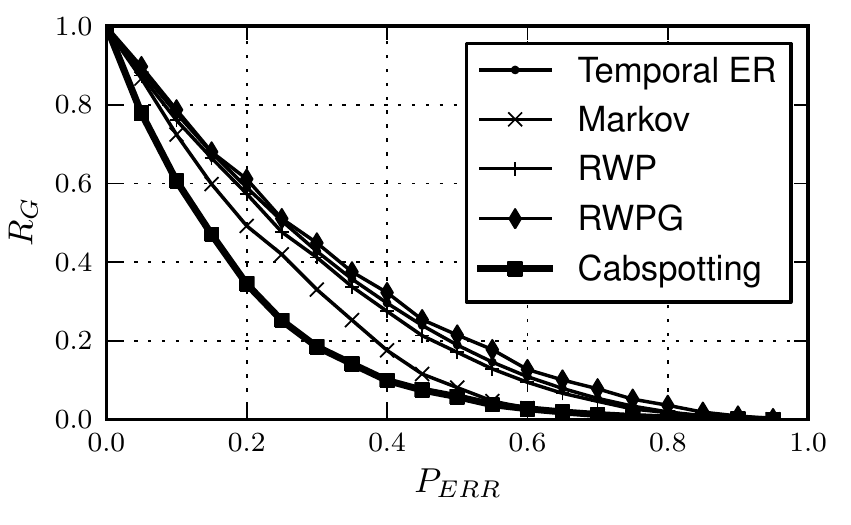}}
  \caption{Comparison between temporal robustness $R_G$ and static robustness as
      a function of probability of error $P_{ERR}$ for the Cabspotting dataset
          (Figure~\ref{fig:cabspotting_robustness_static}).
          The static approach overestimates system robustness.
      Right: Comparison between the temporal robustness $R_G$ of the 
      dataset and random null models
    with the same number of nodes $N$ and $P_{ON}$ of the
    Cabspotting temporal network
          (Figure~\ref{fig:cabspotting_robustness_null}).}
    \label{fig:cabspotting_robustness}
\end{figure}

\subsubsection{Analysis} 
We study the reaction of the Cabspotting temporal network  to random failures
and compare it to our findings on random models. We adopt numerical simulation,
    but since the temporal dynamics of this network is not stationary, we can
    not compare the temporal efficiency $E_G$ before and after a certain error,
    because the two temporal window will likely have already different
    properties. Instead, we fail nodes according to $P_{ERR}$ at the very first
    time step of the temporal sequence of graphs: in this way, we can compare
    the average temporal efficiency over all the time for the original network
    and for the damaged one.
    We adopt a value of $\tau=3600$, which allows us to consider temporal paths
    up to 1 hour, even if such longer paths can not contribute much to temporal
    efficiency.

The first comparison that we show in
Figure~\ref{fig:cabspotting_robustness_static} is
between static robustness and temporal robustness for the Cabspotting temporal
network. In this case static robustness is computed on the static graph obtained
by aggregating all the contacts in the trace and adopting static global
efficiency as performance measure. Since the resulting static graph contains
more than 100,000 edges it is clearly an overestimation of the communication
properties of the real system, as not all these links are continuously available
over time and some paths can not be used due to temporal ordering
constraints. Indeed, static robustness appears much larger than the temporal
counterparts: only temporal robustness is able to capture the realistic
communication capabilities of the system and how they are affected by random
failures. 

Then, we attempt to understand if the various random temporal network models we
have studied can be used to approximate the robustness properties of the real
scenario. For each model, we compute the temporal robustness as a function of
$P_{ERR}$ for a network with the same number of nodes $N$ and the same $P_{ON}$
measured in the Cabspotting temporal network (about $0.005$), using the same
simulation parameters as in the real scenario. As reported in
Figure~\ref{fig:cabspotting_robustness_null}, all
temporal networks present the same trend in network robustness, albeit random
models have higher values of temporal robustness than the real network.
Interestingly, the closest match is the Markov-based temporal model,
while the mobility-based models are closer to the ER model than to the
Cabspotting network, even if this is actually a mobility-based contact
network. However, the assumptions used in mobility models require
homogeneity of space and absolute freedom to move continuously and
independently in a boundless area, while in reality taxis are usually
constrained to move on streets and bridges and they often move together
along the same direction or stop together in a particular place to wait
for customers (i.e., airport or stations). The Markov model, instead, introduces
the type of time correlations that appear to better mimic
the real scenario. In fact, the most important aspect that needs to be captured
is time ordering of events: in random mobility models connections do not follow
 particular time patterns, whereas real traces do (rush
        hour, working hours, human sleeping cycles). Only temporal robustness
can take into account these unique characteristics.

\subsection{Summary of the Findings}
These two results provide evidence that {\em temporal robustness is a
more accurate measure to be used on mobile networks instead of standard static
approaches}.
Therefore, when testing protocols and
applications to be deployed in mobile networks, a temporal study is more
meaningful and should not be substituted by a static approximation.

%% file: Malware.tex
\subsection{Overview}
Smartphones are not only ubiquitous, but also an essential part of life for many people who carry such devices through their daily routine.  It comes at no surprise then that recent studies have shown that the mobility of such devices mimic that of their owners' schedule~\cite{eagle_reality_2006,wang_understandingspreading_2009}. 
This fact constitutes an opportunity for devising efficient protocols and applications, but it also represents an increasing security risk: as with biological viruses that can spread from person to person, mobile phone viruses can also leverage the same social contact patterns to propagate via short-range wireless radio such as Bluetooth and WiFi.  
For example, when security researchers downloaded \textit{Cabir}~\cite{cabir} -- the first proof-of-concept piece of mobile malware -- for analysis, they soon discovered the full risk potential of the mobile worm as it broke loose, replicating from the test device to external mobile phones. 
This event prompted the need for specially radio shielded rooms to securely test such malicious code~\cite{f-secure_2005}.

\jt{removed examples of mobile viruses to save space and wrong audience}

Unlike desktop computers mobile malware can spread through both short-range radio (i.e., Bluetooth and WiFi) and long-range communication (SMS, MMS and email)~\cite{leavitt_mobile_2005}.  Long-range malicious traffic can potentially be contained by the network operator by scanning every message against a database of known malware or spam~\cite{van_ruitenbeek_quantifyingeffectiveness_2007}, 
however, short-range propagation might fall under the radar of centralised service providers: effective schemes to defend against short-range mobile malware spreading are necessary.
In addition, while a global patching of the devices through cellular connectivity is the natural solution and is in theory possible, in practice, due to associated costs and resource consumption, this is not ideal. For example, there are potential constraints with respect to the cellular network capacity and server bandwidth (with respect to the latter, similar issues have also investigated for software updates distribution in the Internet, see for example~\cite{GKC06:planet}).  

\subsection{Temporal Centrality Metrics for Malware}
Being highly correlated with human contacts, understanding how such malware propagates requires an accurate analysis of the underlying time-varying network of contacts amongst individuals.
State-of-the-art solutions on mobile malware containment have ignored two important temporal properties: firstly, the time order, frequency and duration of contacts; and secondly, the time of day a malicious message starts to spread and the delay of a patch~\cite{zhu_social_2009,zyba_defending_2009}. Instead, we argue that the temporal dimension is of key importance in devising effective solutions to this problem. 

With this in mind, the focus of this study is to investigate the effectiveness of two containment strategies based on targetting key nodes, taking into account these temporal characteristics. 
We firstly investigate a traditional strategy, inspired by studies on error and attack tolerance of networks~\cite{albert_error_2000}, exploiting static and a time-aware enhanced version of \textit{betweenness centrality} which provide the best measure of how nodes that mediate or bridge the most communication flows.
According to this strategy the nodes that act as mediators are patched to \textit{block} the path of a malicious message. However, due to temporal clustering and alternative temporal paths, in most cases, such strategies merely \textit{slow} the malware and does not \textit{stop} it. 
In other words, a scheme based solely on immunisation of key nodes is not sufficient, instead {\em quick spreading} of the patch is necessary for most networks.
We propose a solution based on local \textit{spreading} of patches through Bluetooth, i.e., exploiting the same mechanism used by the malware itself. The key issue in this approach is to select the right nodes as starting points of the patching process.
using \textit{temporal closeness centrality} which ranks nodes by the speed at which they can disseminate a message to all other nodes in the network. We show that this strategy can reduce the cellular network resource consumption and associated costs, achieving at the same time a complete containment of the malware in a limited amount of time.

\subsection{Evaluation}
\begin{table}[t!]
\begin{center}
\begin{scriptsize}
\begin{tabular}{|c||c|c|c|}
 \hline
 & CAMBRIDGE  & INFOCOM  & MIT  \\
 \hline
 N  & 18 & 78 & 100 \\
Duration &  10 Days  & 5 days & 280 days \\
Contacts (avg. per day) & 1927 & 25796 & 231 \\
  Scanning Rate     &   30 sec  &  2 min   &  5 min   \\
\hline 
\end{tabular} 
\end{scriptsize}
\end{center}
\vspace{-10pt}
\caption{Experimental Datasets} \label{tbl:datasets}
\vspace{-14pt}
\end{table}


\subsection{Simulation Setup}

\jt{made table smaller}To evaluate the time-aware mobile malware containment schemes, three traces of real mobile device contacts carried by humans are used: Bluetooth traces of researchers at the University of Cambridge, Computer Laboratory, as part of an emotion sensing experiment \cite{rachuri_emotionsense_mobile_2010}; 
Bluetooth traces of participants at the 2006 INFOCOM conference \cite{cambridge-haggle-2009-05-29}; 
and campus Bluetooth traces of students and staff at MIT \cite{eagle_reality_2006}.  
We shall refer to these as CAMBRIDGE, INFOCOM, MIT, respectively.  Table \ref{tbl:datasets} describes the characteristics of each set of traces.  
All three datasets were constructed from mobile device co-location where participants were given Bluetooth enabled mobile devices to carry around.  When two devices come into communication range of the Bluetooth radio, the device logs the colocation with the other device. 
For the CAMBRIDGE dataset, all 10 days are used as part of the evaluation.
For the INFOCOM dataset, since devices were not handed out to participants until late afternoon during the first day, only the last 4 days are used.
For the MIT dataset, we show results for the first two weeks of the Fall semester representing a typical fortnight of activity.


The top $N_p$ devices are chosen according to the calculated temporal betweenness or temporal closeness centrality ranking from the temporal graph ${\cal G}^w_t(t_p,t_{max})$, where $w$ is set to the finest window granularity, corresponding to the scanning rate of the devices in each dataset (for example, 30 second windows for CAMBRIDGE); and $h$ is set to 1, since higher values of $h$ lead to similar performance of the containment schemes.
The $N_m$ nodes that are initially infected with malicious messages are chosen uniformly randomly. The results are obtained by averaging over 100 runs for each $N_p$.
The static centralities from the static aggregated graph over the time interval $[t_p,t_{max}]$ are also calculated for comparison.

Our evaluation is based on the following assumptions: firstly, when a node receives a patch message, it is immunised for the rest of the simulation (i.e., we assume that the malware does not mutate over time); secondly, there is always a successful file transfer between devices (errors in transmission can be taken into consideration in the assessment of the contention scheme without changing significantly the results of our work, assuming random transmission failures); thirdly, an attacker chooses nodes at random; and finally, we have no knowledge of which devices are compromised (otherwise the best scheme is to patch those devices immediately).
%
%
%
%
\begin{figure}[t!]
\vspace{-5pt}
	\begin{minipage}[h!]{0.49\linewidth} 
		\centering
		\includegraphics[scale=0.16, trim = 0mm 0mm 0mm 0mm, clip]{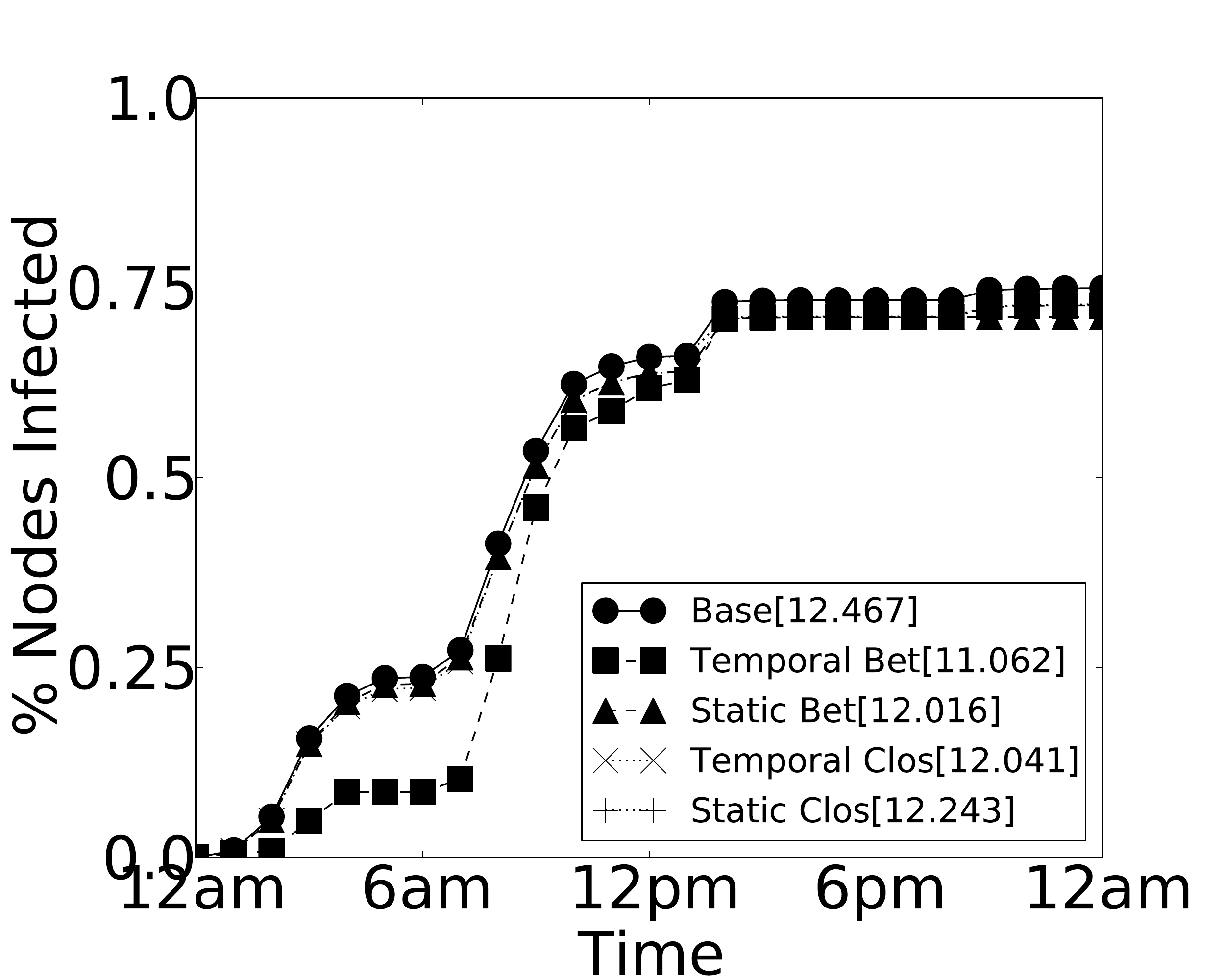}
	\end{minipage}
	\begin{minipage}[h!]{0.49\linewidth}
		\centering
		\includegraphics[scale=0.16, trim = 0mm 0mm 0mm 0mm, clip]{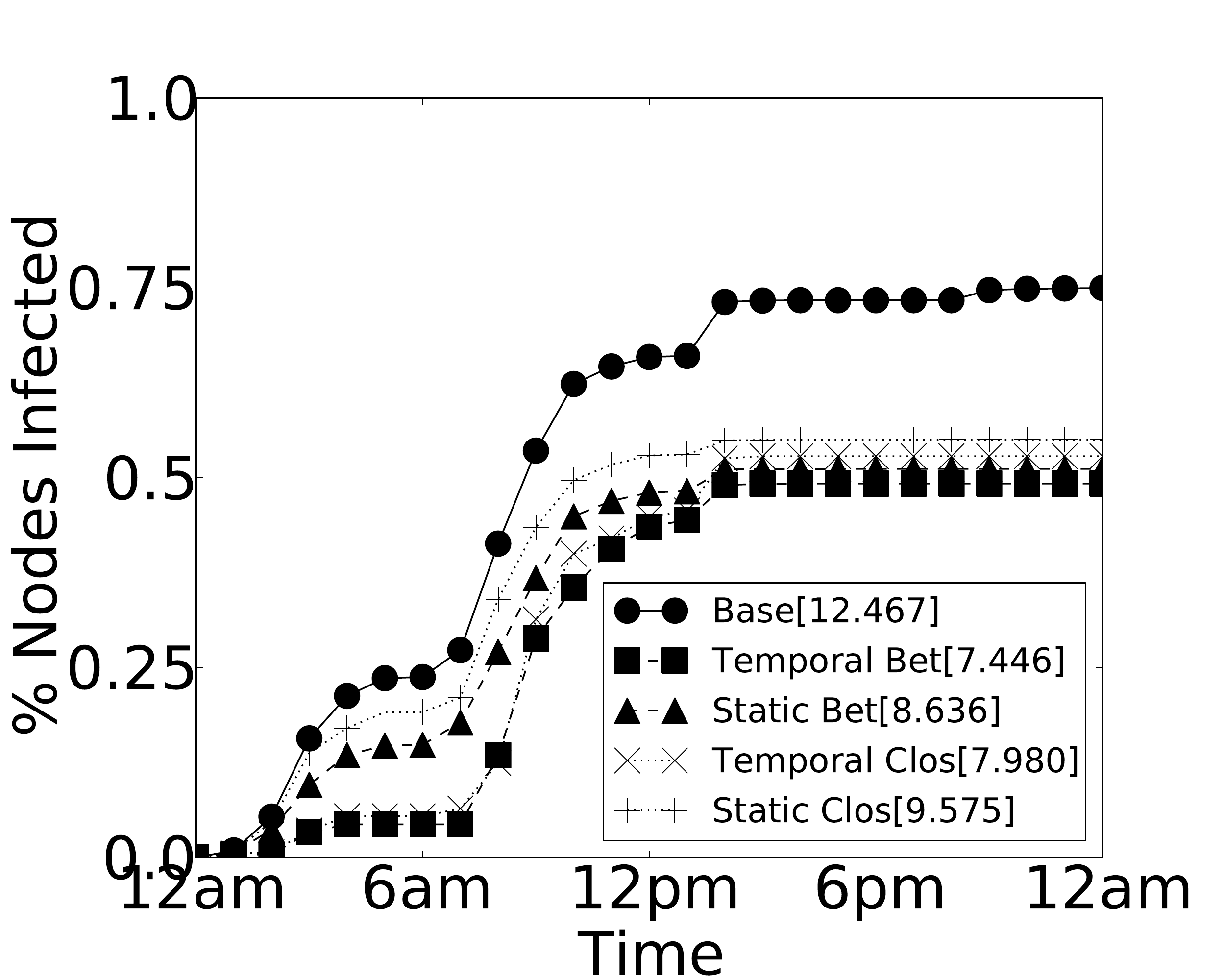}
	\end{minipage}
%
	\begin{minipage}[h!]{0.49\linewidth} 
		\centering
		\includegraphics[scale=0.16, trim = 0mm 0mm 0mm 10mm, clip]{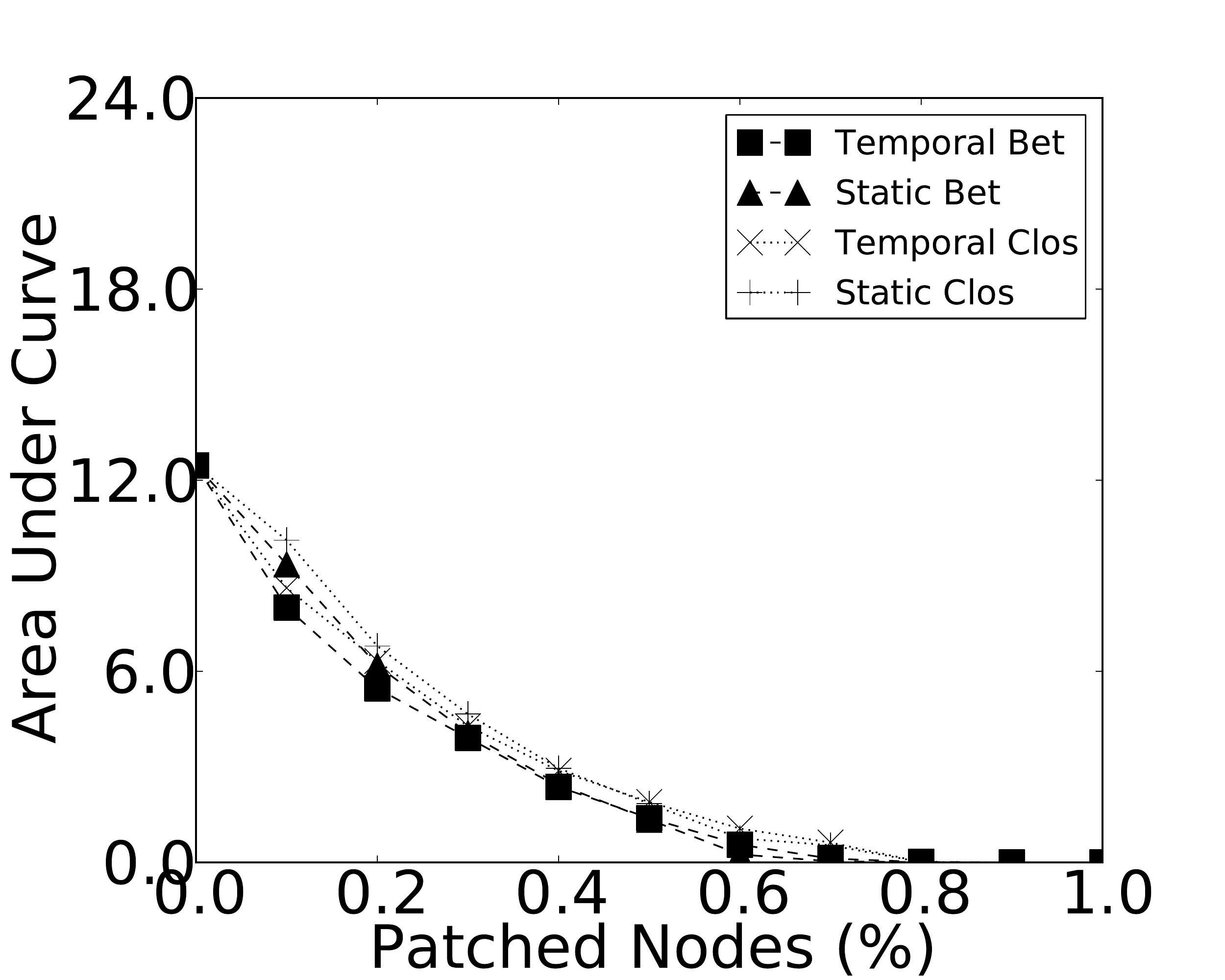}
	\end{minipage}
	\begin{minipage}[h!]{0.49\linewidth}
		\centering
		\includegraphics[scale=0.16, trim = 0mm 0mm 0mm 10mm, clip]{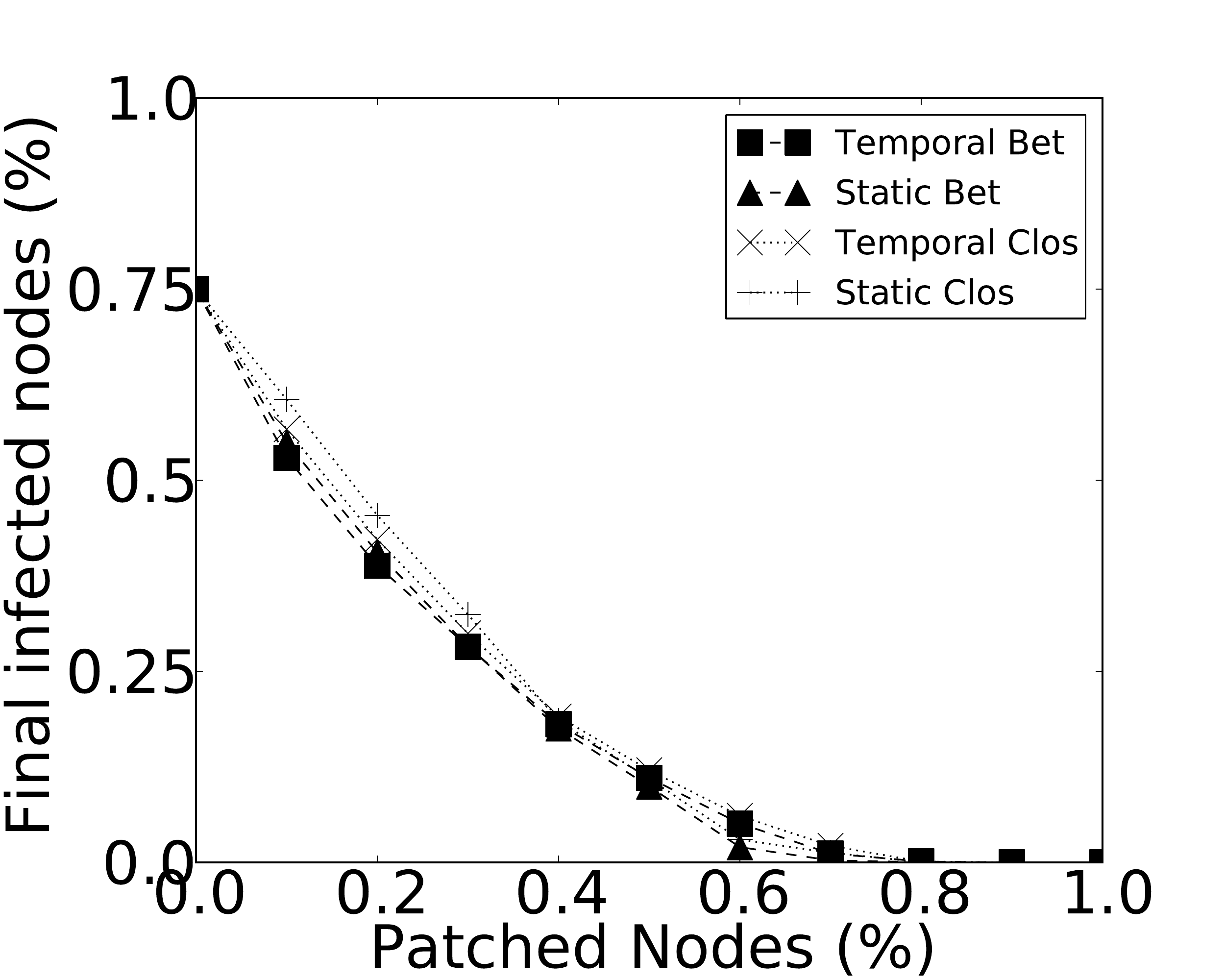}
	\end{minipage}
	\caption{INFOCOM day 4: Immunising 1 node (top left) \& 10 source nodes (top right). Area under curves shown in the legend.  Area (bottom left) and final \% infected nodes (bottom right), as we increase the \% nodes immunised (x-axis).}
	\label{fig:individual_nodes}
	\vspace{-11pt}
\end{figure}
\subsection{Non-Effectiveness of Betweenness based Patching}

%
%
%
Starting from the results of the analysis of the effects time of day has on message spreading, we now evaluate the \textit{best case} scenario for the containment scheme based on patching nodes (without spreading the patch) and we show that this is highly inefficient since it requires a very large number of nodes to be patched via the cellular network 
\jt{deleted "in order"}
to be effective.

Using Day 4 of the INFOCOM trace for this example, a piece of malware is started at the beginning of the day ($t_m$=12am) and the device(s) are patched at the same time ($t_p$=12am).  
This is the best case scenario for two reasons: first, the temporal graph in the morning is characterised by low temporal efficiency since there are very few contacts, therefore, the malware spreads slowly 
; secondly, devices that are immunised immediately have the best chance of blocking malware spreading routes.

Figure~\ref{fig:individual_nodes} shows the ratio of compromised devices across time when the top 1 (top left panel) and top 10 (top right panel) devices are patched after being selected using betweenness and closeness.
As we can see, temporal betweenness 
initially perform better than static betweenness and both temporal and static closeness (quantified by the difference in the area under each curve, shown in the legend). However, by 7am we observe a steep rise in the number of compromised devices and by the end of the day, all curves converge to the same point.  We also note that {\em in both cases it is not possible to totally contain the malware, suggesting that more devices need to be patched}.  Taking a broader view, Figure~\ref{fig:individual_nodes} shows the area under the curve (bottom left) and final ratio of nodes infected (bottom right) as we increase the number of patched devices.  
\jt{replaced "As we can see"}
Clearly, even when the malware is started at the slowest time of day for communication, we still need to patch 80\% of the devices before we can completely stop the malware from spreading; this can be considered an impractically high number of devices to patch.  Similar high percentages are also required in the MIT trace with a minimum of 45\% patched nodes.

\label{sec:individual}
\subsection{Effectiveness of Closeness based Patching (Worst Case Scenario)}
\label{sec:competitive}
\begin{figure}[t!]
	\begin{minipage}[h!]{0.49\linewidth} 
		\centering
		\includegraphics[scale=0.3, trim = 0mm 0mm 0mm 0mm, clip]{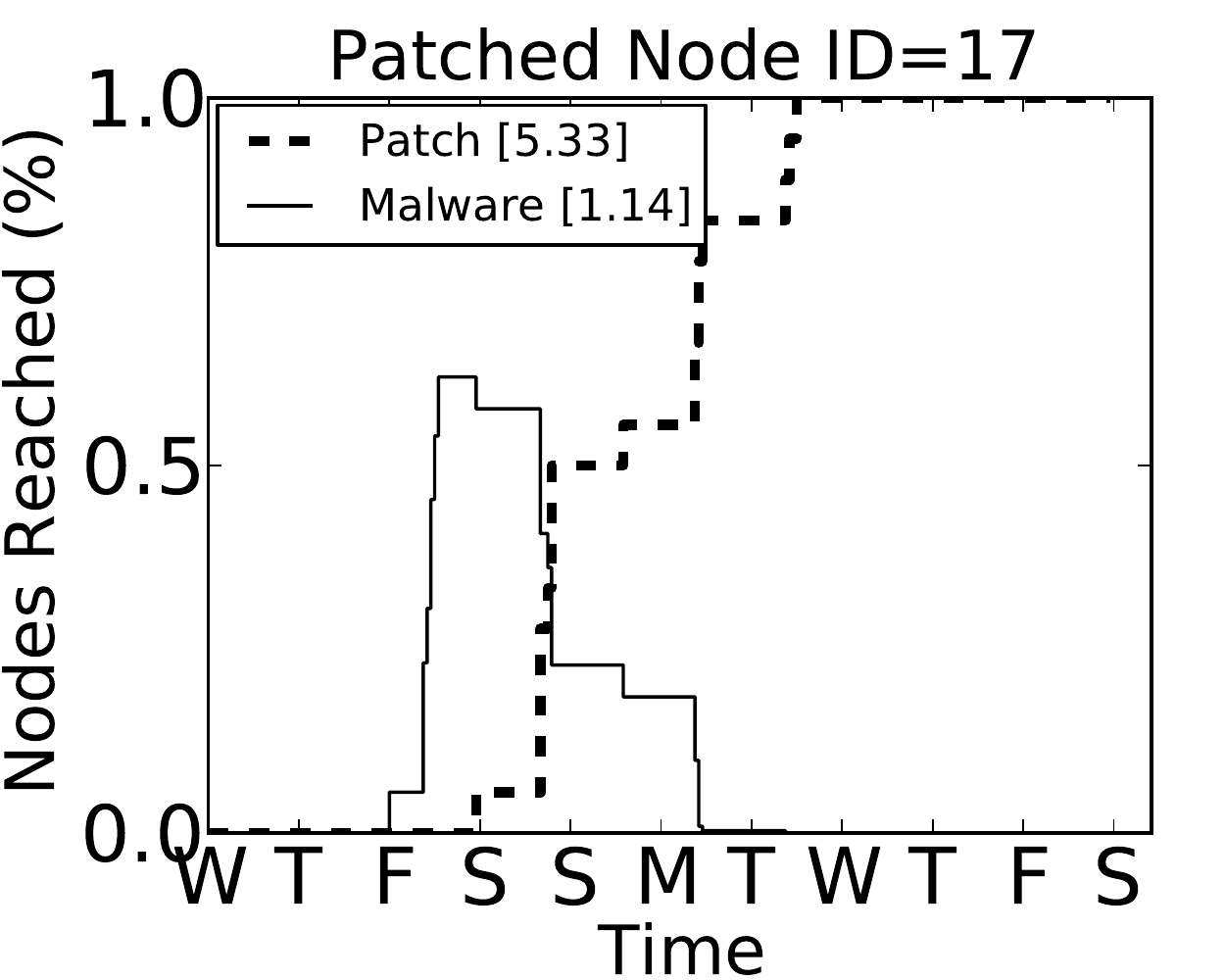}
	\end{minipage}
	\begin{minipage}[h!]{0.49\linewidth}
		\centering
		\includegraphics[scale=0.3, trim = 9mm 0mm 0mm 0mm, clip]{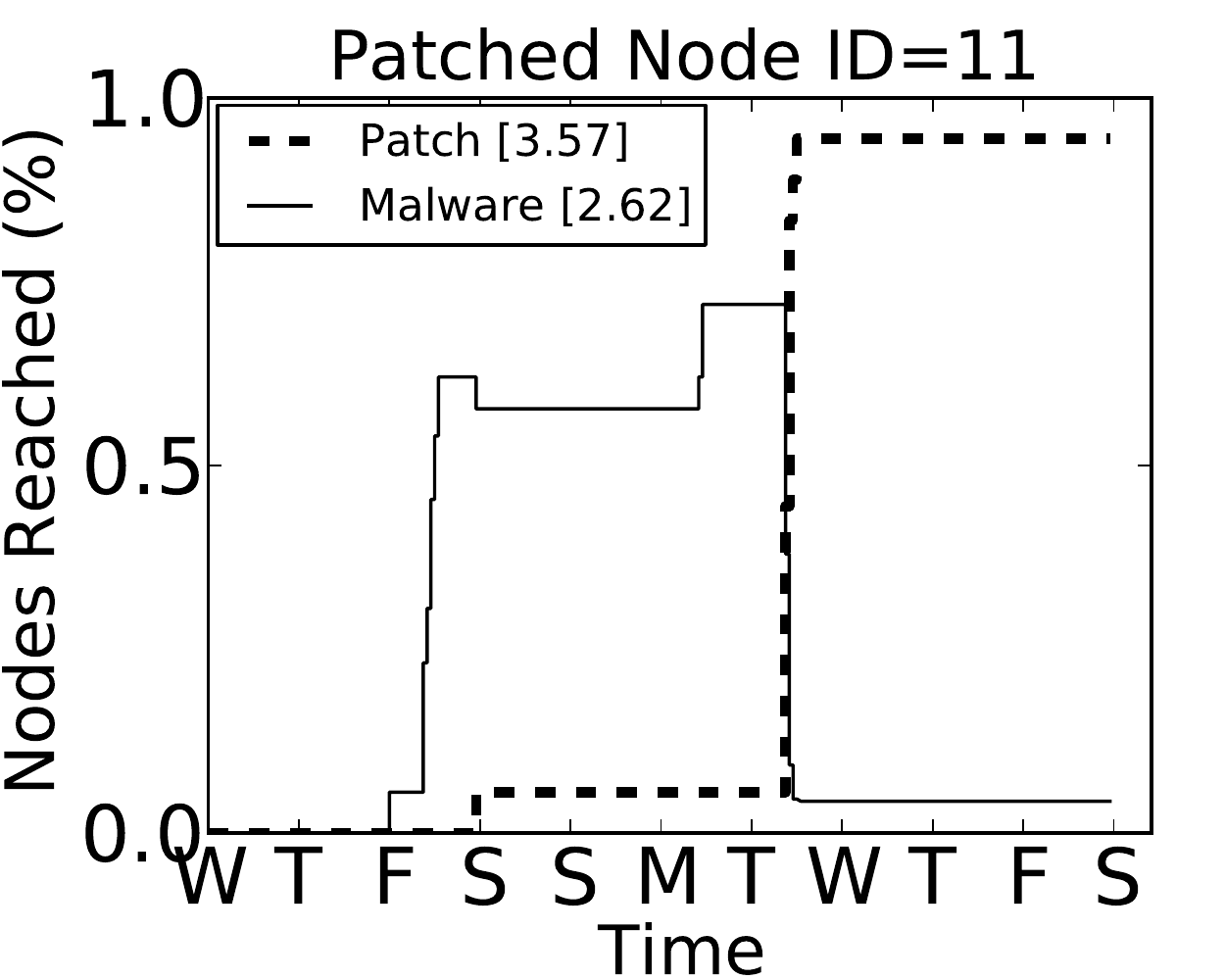}
	\end{minipage}
\vspace{-8pt}
	\caption{CAMBRIDGE [$t_m$=Fri 12am, $t_p$=Sat 12am] delivery rate (y-axis) starting a mobile worm from single node 
with best case patching node (left) and worse case patching node (right) shown. Area under the curve shown in the legend.}\label{fig:ex1}
\vspace{-10pt}
\end{figure}
Since the betweenness based containment scheme is not effective, we now evaluate the closeness based \textit{spreading} scheme with the aim of disseminating a patch message throughout the network more quickly than a malicious message.   For brevity, we do not present results on spreading based on temporal betweeness centrality since it is intuitive that this metric is not designed to quantify the speed of the patching dissemination process and, for this reason, it leads to poorer results.
We start our analysis by examining a \textit{worst case} scenario using the CAMBRIDGE dataset: a researcher receives a malicious message on their device in the early hours of Friday morning ($t_m$=Fri 12am) and the malicious program replicates itself to any devices it meets during the day. A patch message is started a day later to try and patch all the compromised devices ($t_p$=Sat 12am).  
This can be considered as the a worst case since there are more interactions and hence more opportunity for malware to spread during the day and the patch is delayed until a day later.


Figure~\ref{fig:ex1} shows the spreading rate for the malicious message versus the best (left) and worst device (right) to start the patching message. These results were obtained by running simulations considering every single device as a starting point of the patching process, and then ranking them based on three \textit{performance metrics}:
\begin{itemize}
\item the area under the curve (AUC), which captures
the behaviour of the infection over time with respect to the number of infected devices\footnote{The AUC is commonly used in epidemiology and medical trials \cite{trials_small_2001}.};
\item the peak number of compromised devices ($I_{max}$);
\item the time in days necessary to achieve total malware containment ($\tau$).
\end{itemize}
Since  the AUC captures both the $I_{max}$ and $\tau$, the best and worst initial devices that were patched were selected using the AUC.  Comparing all three measures, the case related to the selection of the worst device (right panel) is characterised by double AUC (2.62 vs. 1.07); a higher peak in compromised devices $I_{max}$ (68\% vs. 60\%) and by the fact that it is not possible to fully contain the malware in a finite time $\tau$ ($\infty$ vs. 3.3 days). 
\begin{figure}[h!]
		\center
		\begin{tabular}{c c}	
		\includegraphics[scale=0.24, trim = 0mm 0mm 45mm 1mm, clip]{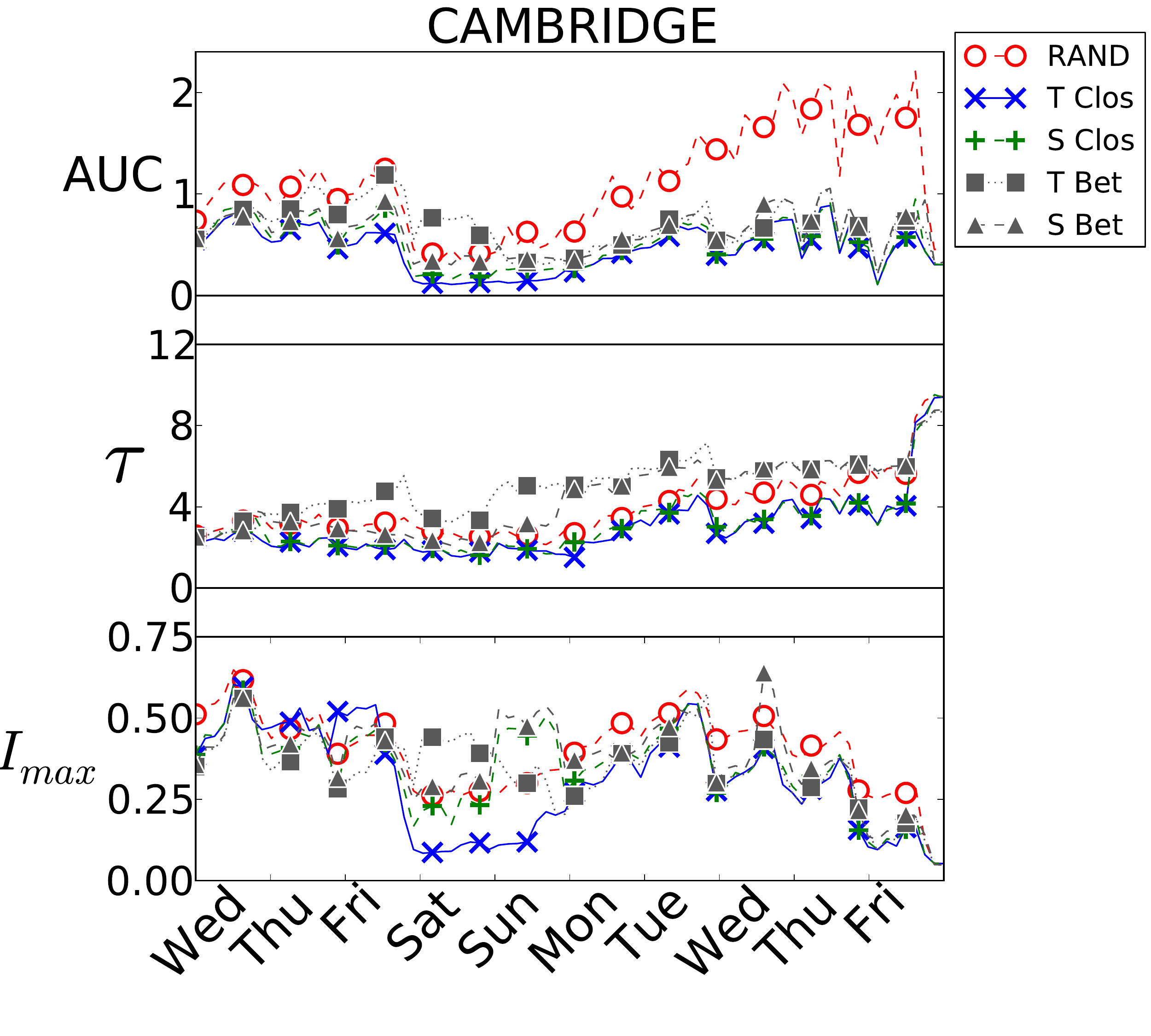}&
		\includegraphics[scale=0.24, trim = 0mm 0mm 45mm 1mm, clip]{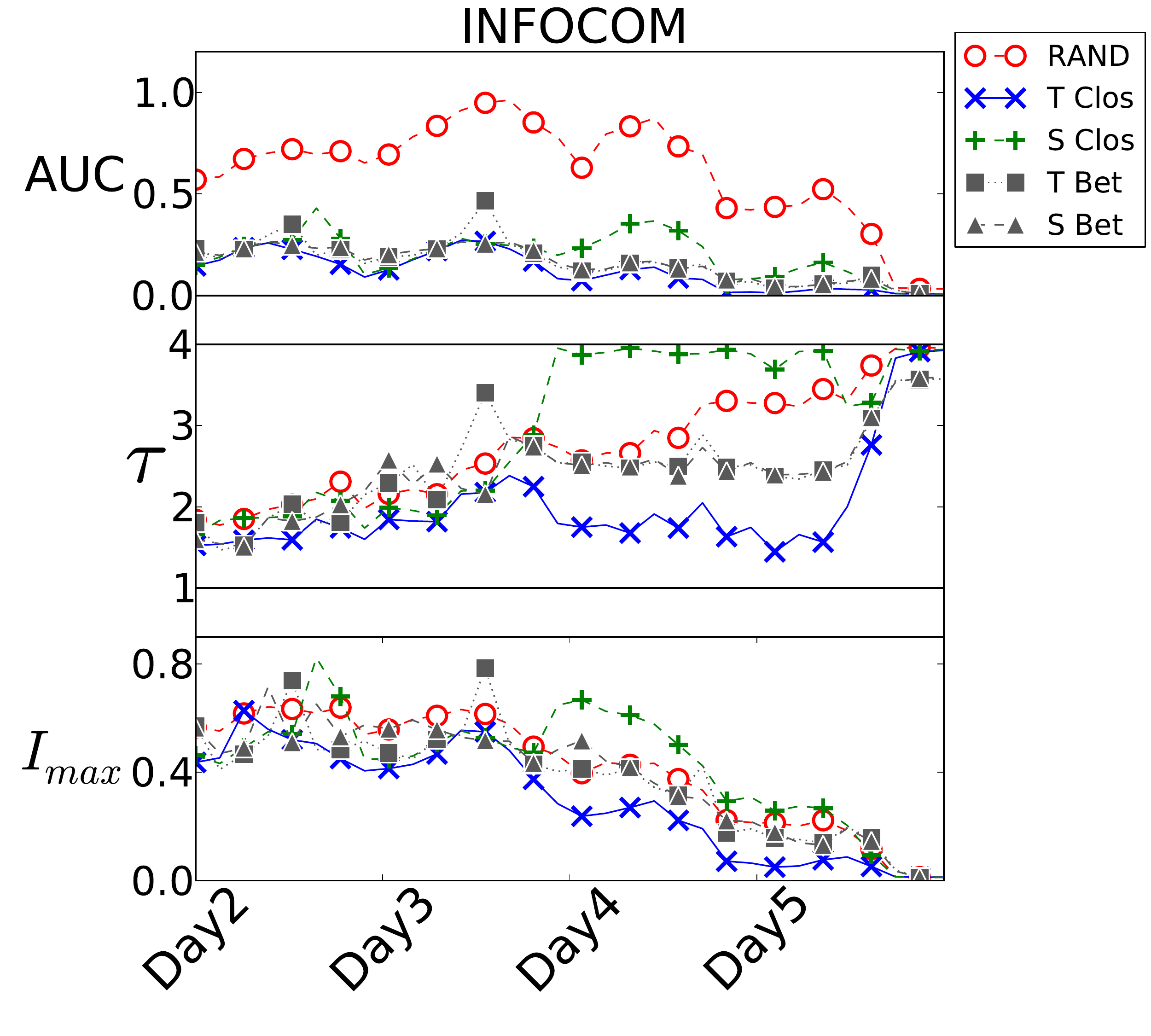}\\
	\multicolumn{2}{c}{\includegraphics[scale=0.24, trim = 0mm 0mm 0mm 1mm, clip]{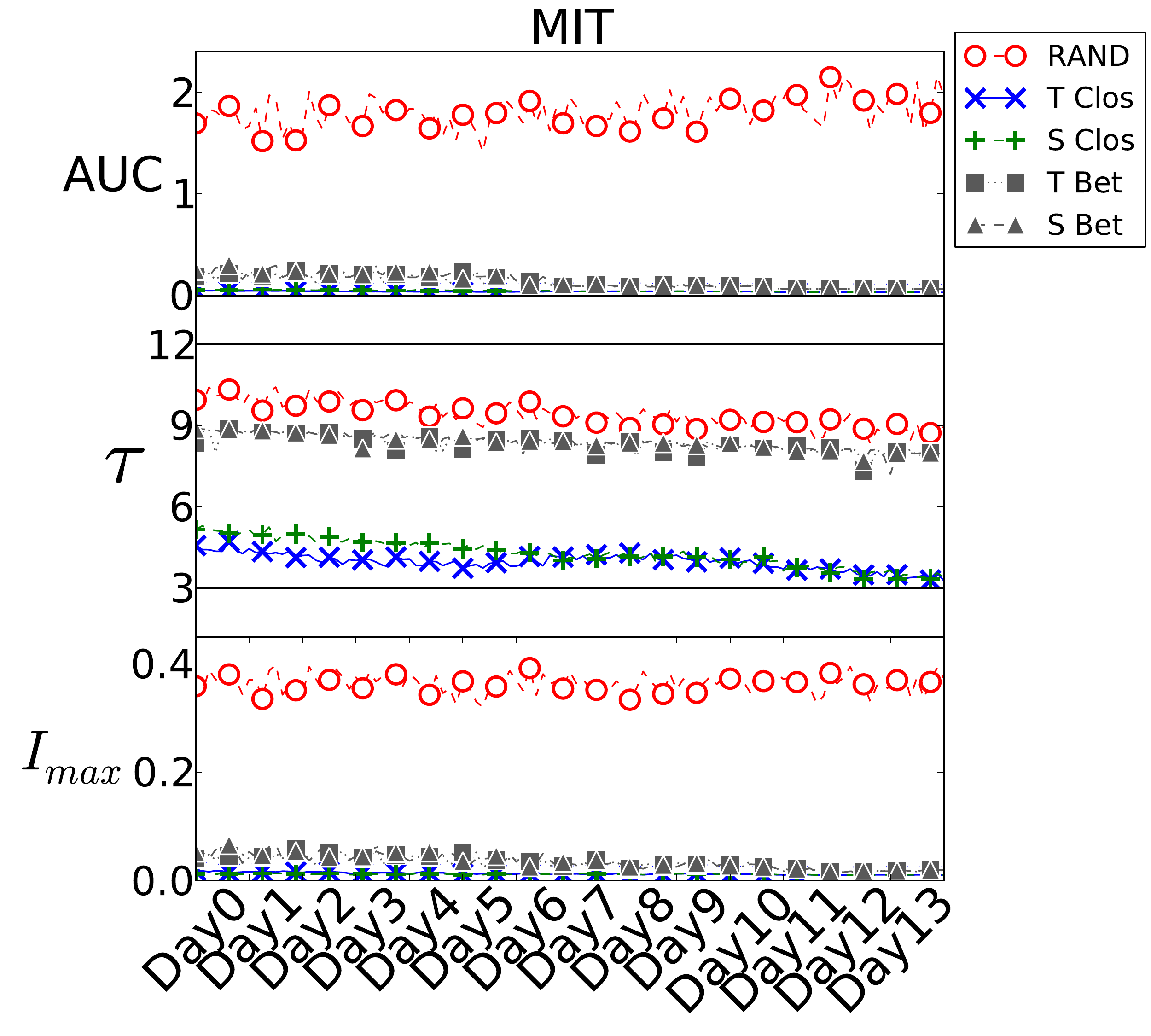}}\\
	\end{tabular}
\vspace{-12pt}	
	\caption{Performance of temporal, static and naive node selection, across different malware start times (x-axis), averaged over all patch delays.}\label{fig:Iall_N1_Pavg}	
\vspace{-12pt}
\end{figure}

\subsection{Sensitivity to Malware Start Time}
%
%



Thus far we have only considered a single malware start time. We now take a broader view and examine the effects of a malicious message starting at different times.  For each dataset the AUC, $I_{max}$ and $\tau$ are exhaustively calculated for different malware start times at hourly intervals and increasing patch delays starting from zero (i.e., patch messages start at the same time as malicious messages) to up to 2 days.  As a baseline, a naive method of randomly selecting patching nodes is also calculated, averaged over 100 runs. Figure~\ref{fig:Iall_N1_Pavg} shows for each dataset the performance metrics as a function of the malware start time $t_m$, averaged over all patch delays.  
In particular, we note that the AUC and the maximum number of infected nodes $I_{max}$ tend to follow the temporal efficiency (strictly related to human circadian rhythms); however, the total time of containment ($\tau$) remains stable across all start times.
These results demonstrate that this time-aware containment scheme is an effective method of quickly containing malware, irrespective of when the malware started.
%
%
%
%
Now analysing the AUC and $I_{max}$, the temporal centrality curve is consistently lower than static and naive methods. Furthermore, static centrality performs worse than the naive method at some points of time; more specifically: 
\begin{itemize}
\vspace{-0.5pt}
\item For the CAMBRIDGE dataset, during the weekend a static method has a higher peak number of compromised devices ($I_{max}$) than the naive method, which shows that a static method is not effective at slowing down the malware from spreading.
\item For the INFOCOM dataset, again $I_{max}$ is higher than the naive method, during days 2 and 4. 
In addition, the AUC curve for a static method peaks with temporal efficiency during days 2, 4 and 5: this means that the malware is not contained effectively in these scenarios.  
Also, the total containment time ($\tau$) is greater than that of the naive method during days 3, 4 and 5. This shows that temporal centrality is more consistently effective for identifying the best nodes to start the patching process, compared to both static and naive methods.
\item Finally, for the MIT dataset, the naive method performs extremely poorly (with high values of UAC, $I_{max}$ and $\tau$ across all malware start times), compared to either a static or temporal method.  
\end{itemize}

\subsection{Summary of the Findings}
This study demonstrates that a temporal analysis of mobile device interactions is better suited to real networks where the topology changes over time.  As we have seen, a traditional strategy of patching high betweenness nodes is not effective when temporal topological information is taking into account for both the information dissemination process and also centrality calculations.  Instead, we propose a strategy that can select the best devices to spread the patch in a \textit{competitive} fashion, using the same opportunistic encounters with other devices utilised by the malware itself.  

%% file: Epidemics.tex
A promising application area of the proposed metrics is indeed epidemics modelling in human networks~\cite{May06:network,KeeRoh07:Modeling}.
Currently, the vast majority of the existing models assume an underlying static network~\cite{LEA03:sexual,Newman10:Networks}. By considering a static network, the \textit{temporal order of appearance} of the links (i.e., the sequence of the contact opportunities) is somehow neglected. This fact might have significant implications on the actual realism of the mathematical model, especially in the case of small populations. Another specific aspect that might have a strong influence on the resulting model is the type of mixing patterns~\cite{New03:mixing} that are present in the population.

The problem of modelling the spreading of infection in a time-varying graph and the definition of vaccination strategies 
given the information related to the network and epidemic dynamics (and the correlation between the two) are open and challenging research areas at the same time. Temporal centrality metrics can be used for example to prioritise the vaccination of individuals involved in an immunisation program. Moreover, temporal metrics can be used in general to study the evolution of a disease over time by providing quantitative measures of the time scale of its spreading considering the sequence of infected individuals (or geographic areas) over time.

In the recent years, some works have been focussed on the interplay between changing topology and the epidemic process taking place over the network. For example, in~\cite{SarKas05:modelling} Saram\"{a}ki and Kaski present a model for studying the spreading of an infectious influenza on a dynamic small-world network, by analysing the effect of a dynamic re-wiring process on a Susceptible-Infective-Recovered-style epidemic model, deriving the equations for the the epidemic threshold and spreading dynamics. In particular, the authors show how the epidemic saturation time scale varies with the size of the network and the initial conditions.
In~\cite{LRLH12:exploiting} the authors present the results of different vaccination strategies by simulating the dynamics of sexual disease spreading in empirical contact sequences of individuals. More specifically, the authors analyse the largest outbreak, the average outbreak sizes, and the relative advantages of the different strategies as a function of the infectivity and the duration of the infective state.

In general, social encounter networks, a typical class of time-varying networks, are attracting an increasing attention in the epidemiology community~\cite{DHRK12:social}. Researchers have been employing RFID and sensor techniques for extracting contact traces in order to accurately reconstruct patterns of interactions among individuals, also with respect to the duration of the contacts (see for example~\cite{CHCDGS07:impact,CVBCP10:Dynamics,SKLLFJ10:highresolution}). 
These models might also be scaled up to study disease outbreaks in cities~\cite{EGAMSTW04:modelling}.

Epidemic models have also been applied to diffusion of ideas, behaviour or lifestyle choices in social networks~\cite{Cen10:spread}, for example in order to study the spreading of obesity~\cite{ChrFow07:spread} or smoking~\cite{ChrFow08:collective}. Another interesting and open area is the characterisation of the spreading of an epidemic and the simultaneous dissemination of information about it that might modify the behaviour of individuals (i.e., the dynamics of the underlying time-varying networks). A work in this direction is~\cite{FGWJ09:spread}.

We believe that the application of the proposed metrics to these fields is indeed very promising and might contribute to increase the accuracy, and therefore, the realism of existing models and to develop new ones allowing researchers to extract valuable information and insights from them. In particular, the centrality metrics presented in this chapter and the accompanying one~\cite{NTMMRL12:graphs} in this book can be used to identify the key spreaders in order to define effective containment and immunisation strategies. For example, vaccination can be based on priorities assigned to the temporal centralities of the individuals in case of human diseases, whereas, as far as computer viruses are concerned, \textit{white worms} used for patching the systems could be distributed starting from the nodes with the highest temporal closeness centrality.

%% file: Summary.tex
In this chapter we presented some key applications of temporal metrics to various domains such as centrality analysis in a social network, robustness and epidemiology of computer viruses and diseases. We have shown that temporal graph metrics are able to provide information about the structure of the time-varying networks and the dynamics of processes happening over them that is not possible to extract through the classic static metrics and graph representations.
We hope that the case studies presented in this chapter and the discussions of open problems in the field might be considered as a starting point and a source of inspiration for future applications in these and other fields.